\documentclass[twocolumn]{webofc}

\usepackage[varg]{txfonts}   
\usepackage{slashed}
\begin{document}
%
\title{Towards a theory of baryon resonances} 
%
%

\author{\firstname{Ulf-G.} \lastname{Mei{\ss}ner}\inst{1,2,3}\fnsep\thanks{\email{meissner@hiskp.uni-bonn.de}} }

\institute{Helmholtz-Institut f\"ur Strahlen- und Kernphysik, 
         Bethe Center for Theoretical Physics and Center for Science and Thought,\\
         Universit\"at Bonn, D-53115  Bonn, Germany
\and
       Institut  f\"{u}r Kernphysik (Theorie), Institute for Advanced Simulation (IAS-4),
       and J\"ulich Center for Hadron Physics,\\
       Forschungszentrum J\"ulich, D-52425 J\"{u}lich, Germany
\and
      Ivane Javakhishvili Tbilisi State University, 0186 Tbilisi, Georgia
          }

\abstract{%
 In this talk, I discuss methods that allow for a systematic and model-independent calculation of the
 hadron spectrum. These are lattice QCD and/or its corresponding Effective Field Theories. Assorted results are shown
 and I take the opportunity to discuss some misconceptions often found in the literature.
 }
\maketitle
\section{Introduction:  QCD and excited states}
\label{intro}
QCD is a remarkable theory. Although its Lagrangian can be written in a line
\begin{equation}
{\cal L}_{\rm QCD} = -\frac{1}{4} \, G_{\mu\nu}^a G^{\mu\nu, a} +
\sum\limits_{f} \bar q_f (i D\!\!\!/ - {\cal M} ) q_f + \ldots~,
\label{eq:L}
\end{equation}
its fundamental fields, the quarks and gluons, have never been observed in isolation
but only appear as constituents of the strongly interacting particles, hadrons and nuclei.
In Eq.~(\ref{eq:L}), $D_\mu = \partial_\mu - ig A_\mu^a \lambda^a/2$ is the gauge-covariant
derivative, $A_\mu^a$ ($a=1,\ldots,8$) the gluon field, $G_{\mu\nu}^a =\partial_\mu A_\nu^a
- \partial_\nu A_\mu^a - ig [A_\mu^b, A_\nu^c]$ the gluon field strength tensor,  $g$ is the
SU(3) gauge coupling, $q_f$ a quark spinor of flavor $f$ ($f=u,d,s,c,b,t$),
${\cal M}$ is the diagonal quark matrix and the ellipsis stand for gauge-fixing and CP-violating
terms not considered here.
The quarks come  in two types, the  light ($u,d,s$) and heavy ($c,b,t$) quark flavors,
where light and heavy refers to the QCD scale $\Lambda_{\rm QCD} \simeq 210$~MeV
(for  $N_f = 5, \overline{MS}, \mu = 2\, $GeV). In the absence of the quark masses,
$\Lambda_{\rm QCD}$ is the only dimensionful parameter in QCD that is generated by dimensional
transmutation through the running of the strong coupling $\alpha_s = g^2/4\pi$.

The Lagrangian of QCD  allows us to define two special limits, in which the theory can
be analyzed in terms of appropriately formulated effective field theories (EFTs).
In the light quark ($f=u,d,s$) sector, the effective Lagrangian can be written
in terms of left- ($q_L$) and right-handed ($q_R$) quark fields, such that
\begin{equation}
{\cal L}_{\rm QCD} = \bar q_L \, i D\!\!\!/ \, q_L  + \bar q_R \, i D\!\!\!/ \, q_R 
+ {\cal O}(m_f / \Lambda_{\rm QCD})~.
\end{equation}
As can be seen, left-  and right-handed  quarks decouple, which is reflected in the 
chiral symmetry.  It is explicitly broken by the finite but small quark masses $m_f$. 
Furthermore, chiral  symmetry is spontaneously broken, leading to the eight pseudo-Goldstone bosons,
the pions, the kaons and the eta. These are indeed the lightest hadrons. The pertinent EFT is
chiral perturbation theory (CHPT). 

Matters are very different for the heavy $c$ and $b$ quarks, where the leading order
Lagrangian takes the form
\begin{equation}
{\cal L}_{\rm QCD} = \bar Q_f \,  i v \cdot D \, Q_f + {\cal O}(\Lambda_{\rm QCD} / m_f)~,
\end{equation}
with $v$ the four-velocity of the heavy quark and $Q_f$ denotes a quark spinor of
flavor $f$ ($f=c,b$). Note that to leading order, this Lagrangian
is independent of quark spin and flavor, which leads to {\em SU(2) spin} and {\em SU(2) flavor symmetries} 
(HQSS and HQFS, respectively). The pertinent EFT to analyze the consequences is 
heavy quark effective field theory (HQEFT), which comes in different manifestations. Finally,
in  heavy-light systems, where heavy quarks act as matter fields coupled to the light pions,
one can combine CHPT and HQEFT.

There are various reasons to consider {\bf excited states}. First, the spectrum of 
QCD is arguably its least understood feature. This is often phrased in terms of
questions like: Why do we observe almost only $qqq$ and $\bar q q$ states?
What is the nature of the XYZ and other ``exotic'' states? Where are the glueballs
predicted by QCD? Note that I put the word exotic in quotation marks, because this
usually refers to states that can not be described within the (conventional) quark model. However,
it is very obvious, but not accounted for by many, that the quark model is much too 
simple, particular in the sector of the light quarks. E.g. it does not account for  a whole
class of important players in the hadron  spectrum, the so-called hadronic molecules.
These also provide the bridge to nuc\-lear physics, because  hadrons and nuclei are just
different manifestations of structure formation in QCD which are intimately linked
and should be considered together. It is also important that high-precision data for spectrum studies
have been and will be produced with ELSA at Bonn, MAMI at Mainz, CEBAF at Jefferson Lab, the 
LHCb experiment at CERN, the BES\-III experiment at the BEPCII, GlueX at Jefferson Lab and in 
the future with PANDA at FAIR and other labs worldwide. These data clearly pose a challenge for 
any theoretical approach.

In what follows, I discuss theoretical approaches that will eventually unravel the physics
behind the QCD spectrum. To give a wider perspective, I present results for mesons and baryons,
although the center of attention of this workshop  clearly is the excitation spectrum of the
nucleon. Nevertheless, the QCD spectrum should be seen in a broader perspective. To be more precise,
I only consider methods that are\\[-2ex]

\begin{minipage}{6.0cm}
\hspace{0.5cm}\begin{itemize}
\item {\bf model-independent},\\[-2ex]
\item  {\bf can be systematically improved}, and\\[-2ex]
\item  {\bf allow for uncertainty estimates}.\\[-1ex] 
\end{itemize}
\end{minipage}

\noindent
If one of these conditions is not
fulfilled, a given method will not be considered further. In particular, I eschew models here.
So that leaves us with lattice QCD (LQCD) and EFTs or combinations thereof.
LQCD can get ground-states  and some excited
states at (almost) physical pion masses, but the most   distinctive feature of excited states are
{\sl decays}.
These are only captured for very few states in lattice QCD, as discussed later.

This talk consists of a number of lessons. In Sec.~\ref{sec-1} I give an answer to the question:  What is a
resonance? I also discuss briefly one example how resonance positions in the complex energy plane
look and stress the two-pole structure of the $\Lambda(1405)$. Then, in Sec.~\ref{sec-2}, I discuss the case of a 
well-separated resonance in a finite volume, which brings us in contact with the L\"uscher equation. 
This is extended to the multi-channel case in Sec.~\ref{sec-3}. I show in particular how chiral symmetry is
important for the extrapolation into the complex plane, which leads us to the two-pole structure of the
$D_0^\star(2400)$ from the analysis of high-precision lattice QCD data on the coupled channel 
$D\pi, D\eta, D_s\bar{K}$ scattering with isospin $I=1/2$. I also point out that this two-pole structure
is more common then so far believed. In Sec.~\ref{sec-4}, I discuss hadronic molecules, which are
important players in the hadron spectrum. I address in particular the issues  how to distinguish them from
compact multi-quark states and the calculation of the production cross section in high-energy proton-(anti-)proton
collisions. Then, in Sec.~\ref{sec-5} I turn to the calculation of the width of the two lowest-lying baryon
resonances, the $\Delta(1232)$ and the Roper $N^*(1440)$ using covariant baryon chiral perturbation theory 
and the complex-mass scheme. Sec.~\ref{sec-6} contains a few remarks on the pion cloud of the nucleon and
other hadrons, also often encountered at this workshop. I end with a short summary in terms of take-home messages.

\begin{figure}[t]
\centering
\includegraphics[width=0.40\textwidth]{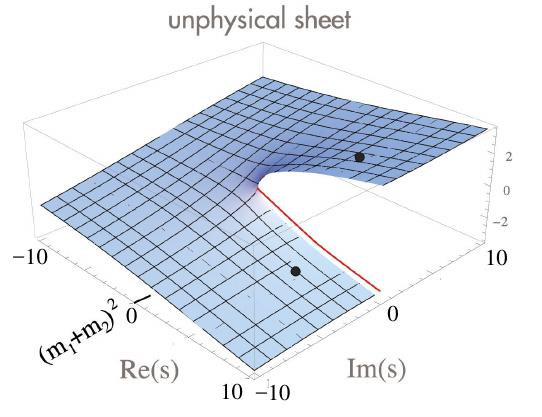}~~~~~
\caption{The imaginary part of a single-channel amplitude in the
  presence of a resonance. The solid dots indicate the allowed positions
  for resonance poles. Figure from~\cite{Guo:2017jvc}.}
\label{fig:resosheet}
\vspace{-4mm}
\end{figure}

\section{Lesson 1: What is a resonance?}
\label{sec-1}

In the old times, resonances were searched for by looking at bumps in scattering  cross sections.
But as Moorhouse stated so eloquently: ``Not every bump is a resonance and not every resonance
is a bump'' \cite{Moorhouse}. Now we know that resonances have {\bf complex} properties, like
their mass and width, their photo-couplings, etc.\ . In particular, these intrinsic properties
do not depend on the experiment or theory (model). Most importantly, resonances correspond to
{\bf S-matrix poles on unphysical Riemann sheets}, as depicted in Fig.~\ref{fig:resosheet}.

\begin{figure}[t]
\centering
\includegraphics[width=0.48\textwidth]{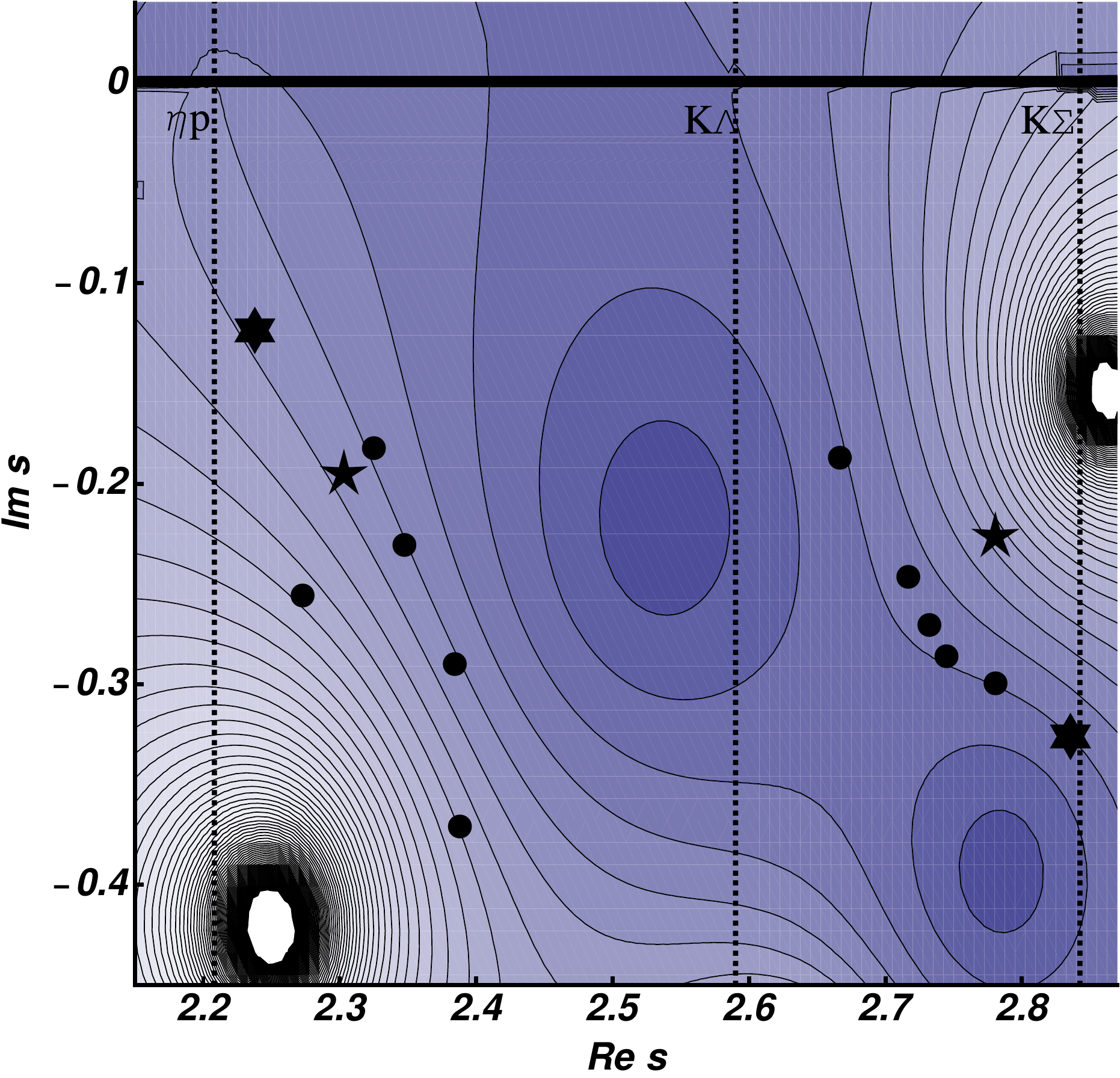}
\caption{Modulus of the analytic continuation of the 
  $J^P = 1/2^-$ $\pi N$ scattering amplitude
  into the complex $s$-plane. Shown is the (- - - + + +) Riemann sheet. The two poles
  at $\sqrt{s}=(1.51-i0.14)$~GeV and $\sqrt{s}=(1.69-i0.05)$, corresponding
  to the $S_{11}(1535)$ and the $S_{11}(1650)$, respectively, are
  clearly visible. The stars and black dots refer to other determinations
  as listed in~\cite{Bruns:2010sv}. For the definitions of the sheets, see e.g. Ref.~~\cite{Cieply:2016jby}.
}
\label{fig:sheet6}
\vspace{-4mm}
\end{figure}

\begin{figure}[t]
\centering
\includegraphics[width=0.40\textwidth]{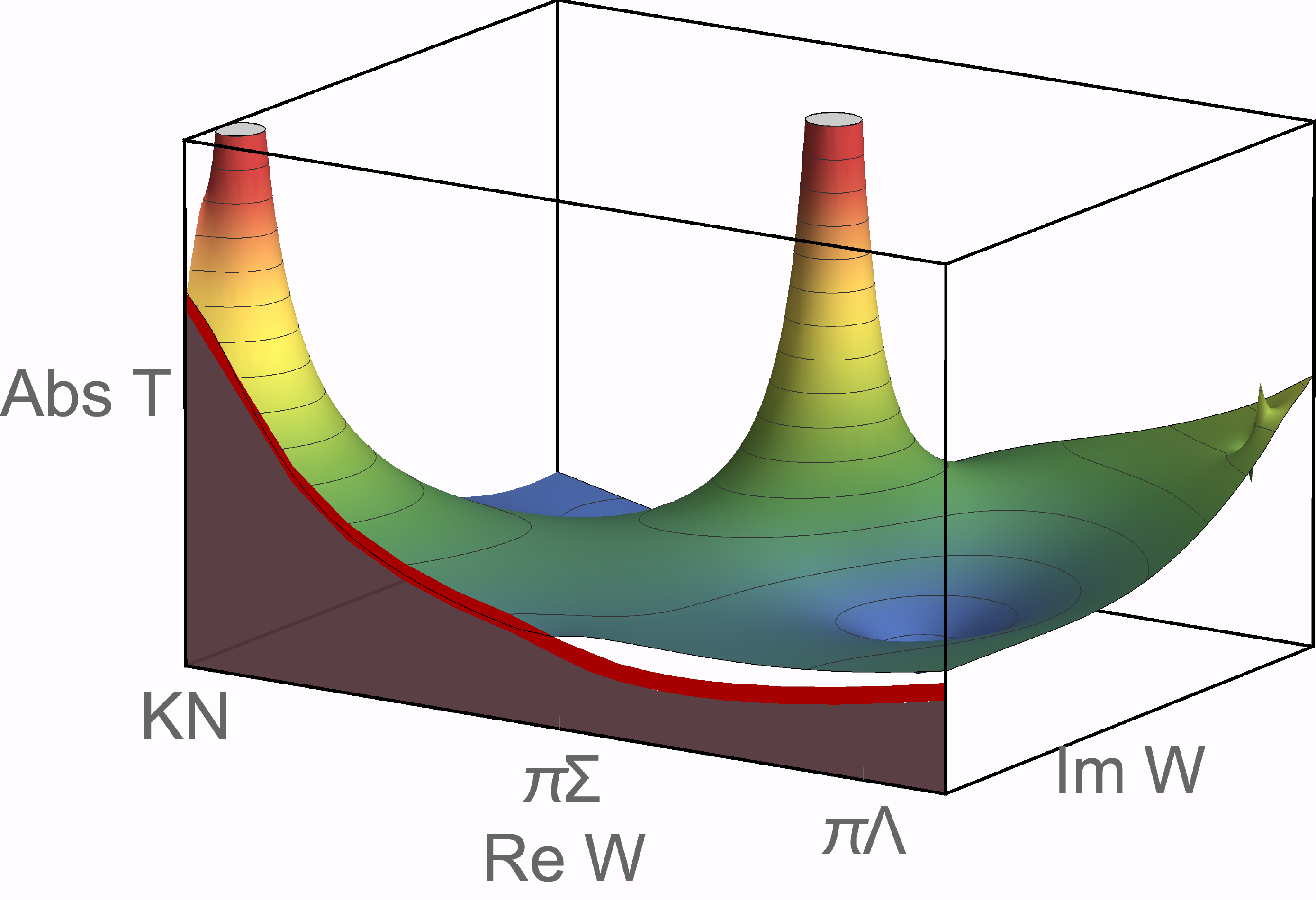}
\caption{The S-wave  amplitude  $f_{0+}$ for $K^-p\to K^-p$ for $I=0$ in the complex energy
plane clearly showing the two poles in the region of the $\Lambda(1405)$. The red band indicates
that the (+ + - - - - + + + +) Riemann sheet is only connected to the real energy axis between the $\bar KN$ 
and the $\pi \Sigma$ thresholds. For the definitions of the sheets, see e.g. Ref.~~\cite{Cieply:2016jby}.
Figure courtesy of Maxim Mai.
}
\label{fig:L1405}
\vspace{-4mm}
\end{figure}

This is the most basic and {\em the only acceptable definition of a resonance} (with very
few exception of well-isolated, single channel cases). So if somebody supplies you
with the resonances parameters and has not looked for the corresponding pole in the
complex plane, these numbers should be considered with suspicion or even be discarded.
Especially in cases when a number of coupled channels is involved, the search in the complex energy plane
is essentially the only viable method. A nice example of such a search in the multiple channel
case (here: $\pi^0p$, $\pi^+n$, $p\eta$, $\Lambda K^+$, $\Sigma^0K^+$ and $\Sigma^+K^0$) 
are the two close-by poles corresponding to the two lowest $S_{11}$ resonances
in elastic pion-nucleon scattering with $J^P=1/2^-$ as shown in Fig.~\ref{fig:sheet6}
(from Ref.~\cite{Bruns:2010sv})
Another beautiful example is given by the two-pole structure of the $\Lambda(1405)$, first noted 
in Ref.~\cite{Oller:2000fj}. Here, multiple channels need to be considered, namely $K^- p \to K^-p, \bar{K}^0 n,
\Sigma^0 \pi^0, \Sigma^+\pi^-, \Sigma^- \pi^+, \Lambda \pi^0, \Lambda\eta, \Xi^+K^-, \Xi^0K^0$. Only if
one analyzes the pole structure in the complex energy plan, one finds these two poles, one being close
to the $K^-p$ and the other closer to the $\pi\Sigma$ threshold, see Fig.~\ref{fig:L1405}. 
In fact, this two-pole structure has been verified by various
groups world-wide, for a comparative study see~\cite{Cieply:2016jby} and later discussions.

Let us now consider the lattice, which corresponds to a finite, cubic box (keeping
the time coordinate continuous for the moment). As it is well-known from quantum mechanics, the eigenstates
of any Hamiltonian in a box are discrete energy levels. So does that mean that the excited states
of QCD are not amenable to LQCD? Fortunately, as pointed out by L\"uscher~\cite{Luscher:1986pf}
and others, one can relate the volume dependence of the energy spectrum on the lattice to the continuum
scattering phase shift. In fact, an isolated narrow resonance can be traced back to an avoided
level crossing, see e.g. Ref.~\cite{Wiese:1988qy}.  This, however, is not a
practical method, so in the following sections I will discuss how to proceed in LQCD.

\section{Lesson 2: Well separated resonances}
\label{sec-2}
\begin{figure}[t]
\centering
\includegraphics[width=0.40\textwidth]{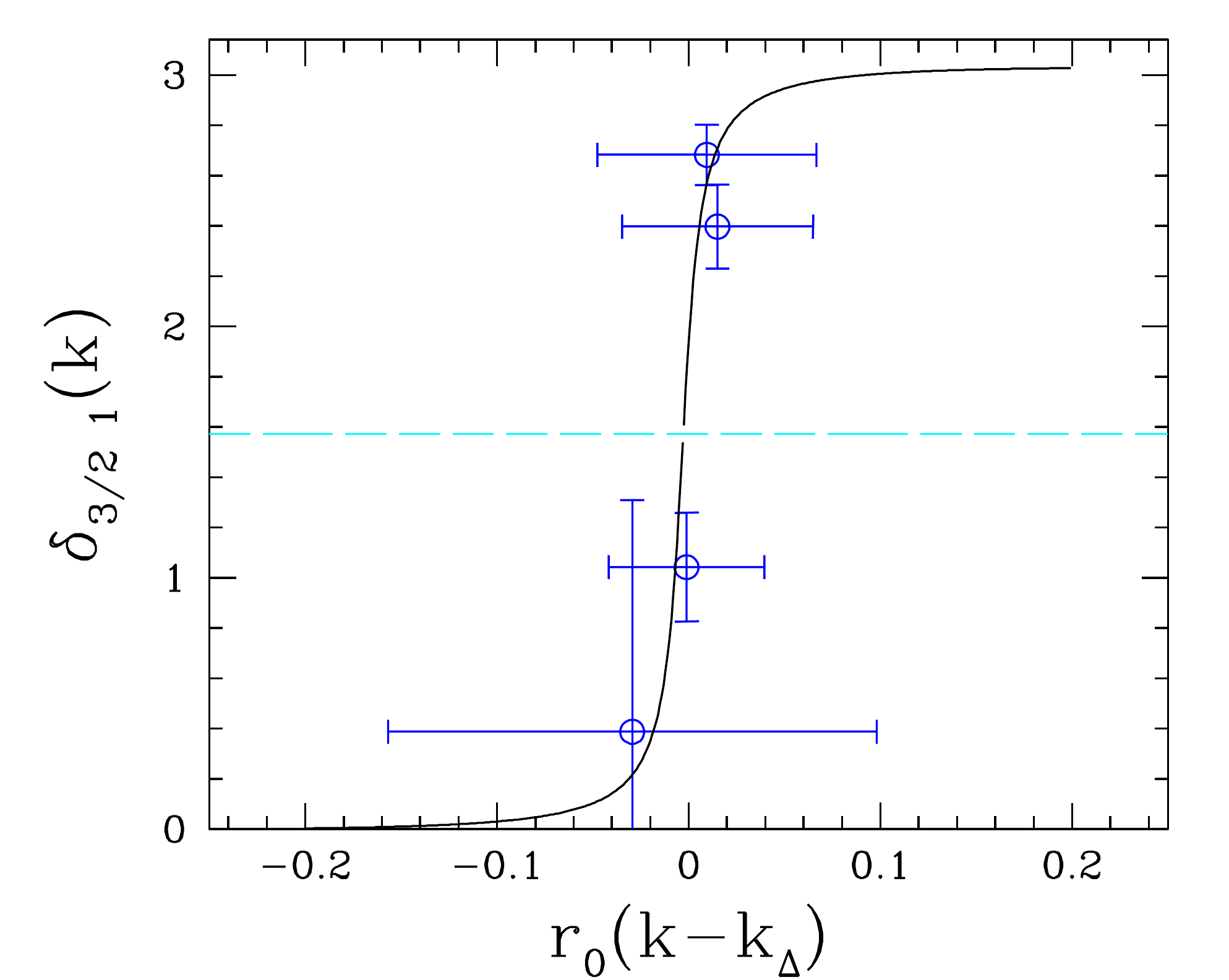}~~~~~
\caption{The $\pi N$ phase shift of the $\Delta$ channel as a function of the
center-of-mass momentum $k$.
The solid line is the phase generated from the physical mass
and width of the $\Delta$-resonance. The dashed horizontal line
indicates the $90^\circ$ crossing of the phase.}
\label{fig:delta}
\vspace{-4mm}
\end{figure}

Well separated resonances are, as already stated, more the exception than the rule.
Still, they provide important benchmarks for any calculation of the spectrum. Let us
consider an isolated resonance in a box.  Before discussing the L\"uscher formalism in three 
space dimensions, let me elaborate on the one-dimensional case as 
an instructive example. Consider the scattering of two particles on a line segment of length $L$ (the lattice size), 
with a finite interaction range $R \ll L$. For simplicity,   assume further an infinite time extent. As expected, outside 
the range of the interaction, the relative motion of the scattering particles should be described by a plane wave. 
Then, the interaction produces a finite phase shift $\delta(k)$ as known from elementary scattering theory. 
Let us  impose periodic boundary conditions in the spatial dimension. Hence, for
a plane wave with momentum $k$, one has
\begin{equation}
\exp(ikL+2i\delta(k)) = \exp(ik0) = 1, 
\end{equation}
which gives 
\begin{equation}
k_n^{}L + 2\delta(k_n^{}) = 2\pi n, 
\quad n \in {\mathbb N},
\end{equation}
for the quantization of the momenta. This result is remarkable, as it relates energies computed on the lattice to 
the continuum phase shift. Given an appropriate (lattice or continuum) dispersion relation which relates the lattice
energy levels $E_n(L)$ to the momentum modes $k_n(L)$, the continuum phase shift at discrete values of the scattering
momentum can be reconstructed from the energy levels on the lattice. Also, the non-interacting limit $\delta(k_n) = 0$ is
recovered for $k_n = 2\pi n/L$. For non-relativistic particles with mass $m$, this method is applicable when 
$m L \gg 1$, so that further corrections are exponentially suppressed. Also, any inelasticity or coupling to other scattering 
channels introduces modifications.  Let me now discuss the case of three space dimensions.
The energy levels for two non-interacting
identical particles of mass $m$ are given by
\begin{equation}
E_n(L) = 2\sqrt{m^2_{}+{\vec k\,}^2}\, ,
~~k_i = \frac{2\pi}{L} n_i ~, ~ n_i \in \mathbb{Z}~.
\end{equation}
Turning on the interaction, we obtain the continuum scattering phase $\delta(k)$ with the
help of L\"uscher's formula,
\begin{eqnarray}
\delta (k) &=&-\phi(q)~ {\rm mod}~\pi~,~~
q = \frac{kL}{2\pi}~,\nonumber\\
\phi(q) &=& -\arctan\frac{\pi^{3/2}q}{{\cal Z}_{00}(1;q^2)}~,\nonumber\\
{\cal Z}_{00} (1;q^2) &=& \frac{1}{\sqrt{4\pi}} \,\sum\limits_{\vec{n} \in \mathbb{Z}^3}
\frac{1}{{\vec n\,}^2 - q^2}~,
\end{eqnarray}  
for the S-wave. Note that the L\"uscher $\zeta$-function ${\cal Z}_{00}$ requires regularization,
which can be done in different ways (a specific example is given in the next section).
Generali\-zations to higher partial waves also exist. If we assume a
resonance with mass $m_R>2m$, we can also use the effective range expansion (ERE),
\begin{equation}
\tan\left( \delta - \displaystyle\frac{\pi}{2}\right) 
              = \displaystyle\frac{E^2 -m_R^2}{m_R\Gamma_R}~,
\end{equation}  
with $m_R ~(\Gamma_R)$ the mass (width) of the resonance. Note, however, that this
can only be used under very special circumstances (well isolated resonance, weak energy
dependence of the background). In such a case, one can measure the phase shift in the
resonance region and fit $m_R, \Gamma_R$, making also use of moving frames as pioneered
in Ref.~\cite{Rummukainen:1995vs}. As an example, I show in Fig.~\ref{fig:delta}
some pioneering (but unpublished) results for the $\Delta(1232)$,  which is a well
separated baryon resonance in the $\pi N$ system, corresponding to the $l=1, I=3/2$~$ \pi N$
phase shift. These results were obtained for pion masses from 160 to 390~MeV in  fairly large
volumes~\cite{BJQCDSF}. More precisely, the calculation used a  $N_f=2$ clover action that consists of the
plaquette gluon action  together with nonperturbatively $O(a)$ improved Wilson (clover) fermions
on $32^3\times 64$ and $40^3\times 64$ lattices at $\beta=5.29,
\kappa=0.13632, a = 0.075\,$fm and on $40^3\times 64$ and $48^3\times 64$ 
lattices at $\beta=5.29, \kappa=0.1364, a = 0.06\,$fm. 
More recent results on the $\Delta$ are given in the talks by
Morningstar~\cite{CM} and by Petschlies~\cite{MP}.
For single channel calculations of elastic pion-nucleon scattering in the
$J^P=1/2^-$ channel, see Refs.~\cite{Lang:2012db,Verduci:2014csa}, and for the
Roper in the $J^P=1/2^+$ channel, see e.g. Ref.~\cite{Leskovec:2018lxb}.

In the meson sector, the genuine case of an isolated resonance is the
$\rho$ that shows up in the $I=J=1$ channel of pion-pion scattering.
Here, the most recent calculation is due to the ETM Collaboration~\cite{Werner:2019hxc},
and I refer to that paper for details.

\section{Lesson 3: Coupled channels/thresholds}
\label{sec-3}

As said before, isolated (well-separated) resonances are the exception, in
most cases we have to deal with coupled channel effects and/or close-by thresholds,
e.g. for the scalar mesons  $f_0(980)$ and $a_0(980)$ or the two poles in the $\Lambda(1405)$
region. There exist various extensions of L\"uscher's approach to cope with coupled channels:
1) a purely quantum mechanical treatment, see e.g.~\cite{He:2005ey}, 2) the formulation
within non-relativistic EFT (NREFT), see e.g. \cite{Lage:2009zv,Bernard:2010fp,Briceno:2012yi,Briceno:2013lba},
3) the use of finite-volume unitarized CHPT, see e.g.~\cite{Doring:2011vk,Doring:2012eu} or 4) a relativistic
version,  see e.g.~\cite{Briceno:2017tce}. The main advantage of the third method is that it is based
on an effective chiral Lagrangian that allows one to relate various processes. This will be exemplified below.
So far, these methods have been mostly applied in the
meson sector, with a few exceptions also for baryons~\cite{Hall:2014uca,Molina:2015uqp}.
Before elaborating on one concrete example, I would like to issue a warning: Be aware of
methods that can mislead you (K-matrix and alike). Some possible pitfalls will also be
discussed in what follows.

\begin{figure}[t]
\centering
\includegraphics[width=0.40\textwidth]{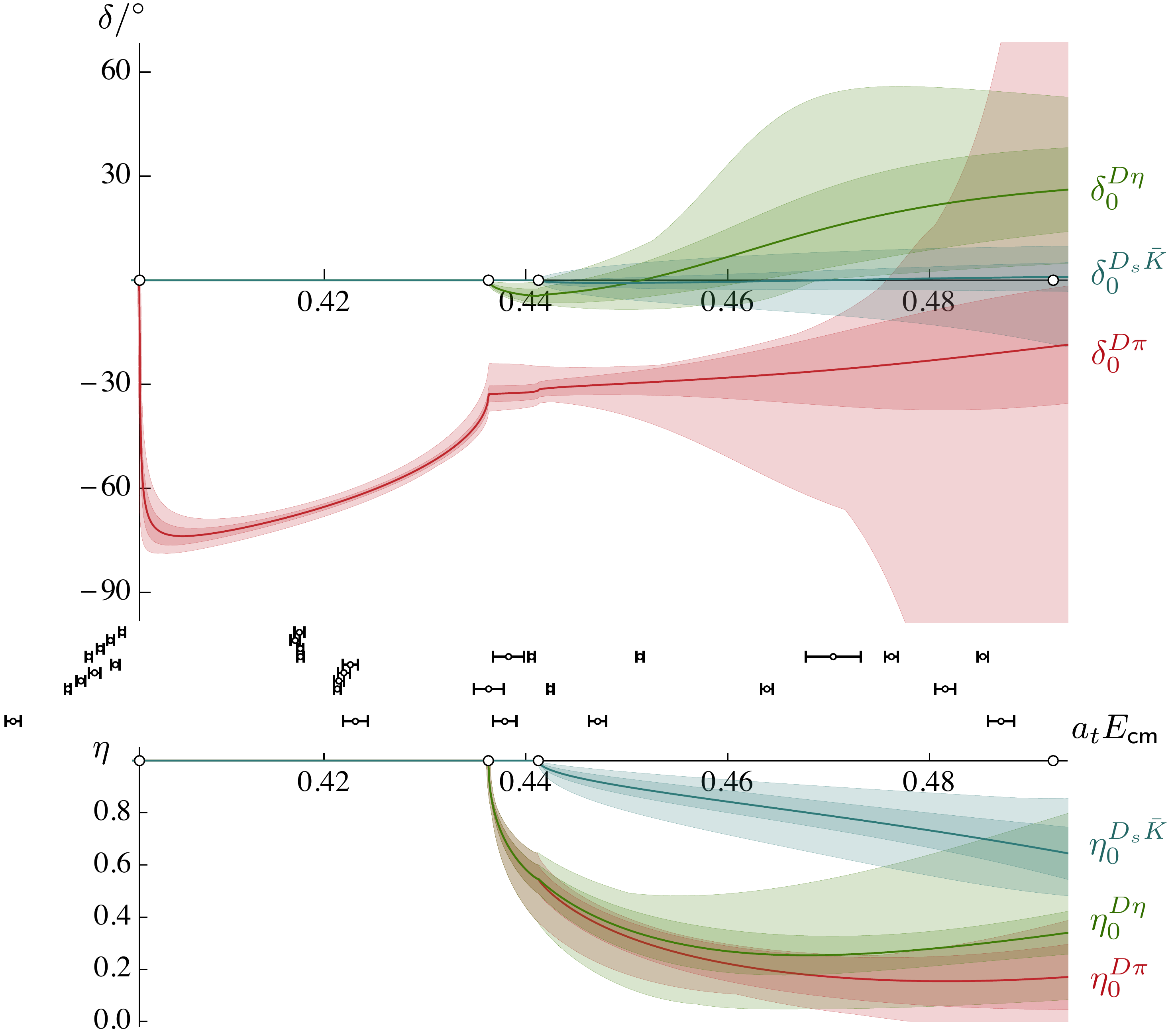}
\caption{The upper (lower)  panel shows the S-wave phase shifts (inelasticities)
  for the $D\pi$ (red), $D\eta$ (green) and $D_{s}\bar{K}$ (blue) channels, where the size of the bands
  incorporates all T-matrix parametrizations of the S-wave. The black points
  show the location of the finite-volume energy levels used to constrain the parametrizations.
  Figure from~\cite{Moir}.
   }
\label{fig:HSC}
\vspace{-4mm}
\end{figure}

The Hadron Spectrum Collaboration (HSC) has investigated the coupled channel $D\pi$, $D\eta$, $D_s\bar{K}$
scattering with isospin $I = 1/2$~\cite{Moir} (for S-, P- and D-waves). They presented results for
three lattice volumes, one spatial lattice spacing $a_s$ and one temporal spacing $a_t$ at $M_\pi
\simeq 390\,$MeV. Furthermore, they used various K-matrix type extrapolations (in fact, up to 11 variants)
to look for poles in the complex plane. The resulting phase shifts and inelasticities are shown in
Fig.~\ref{fig:HSC}. They report one S-wave pole at
$(2275.9 \pm 0.9)$~MeV that is located very closely  to the $D\pi$ threshold. The authors of
Ref.~\cite{Moir} argue that this pole shares similarities with the $D_0^\star (2400)$ of the Particle
Data Group (PDG)~\cite{PDG} (note
that the $D_0^*(2400)$ has been renamed as $D_0^*(2300)$ in the 2019 PDG update).
However, in their T-matrix parametrizations, they ignored the important role of chiral symmetry.
It is long known how to incorporate chiral symmetry in coupled-channel dynamics in the framework
of unitarized CHPT (UCHPT), see e.g. the groundbreaking works in
Refs.~\cite{Kaiser:1995eg,Oset:1997it,Oller:2000fj,Lutz:2001yb}. To make contact to the work of the
Hadron Spectrum Collaboration, let us consider the case of Goldstone bosons scattering off $D$-mesons
($D \phi$ scattering). The corresponding T-matrix in UCHPT takes the form
\begin{equation}
T^{-1}(s) = V^{-1}(s) - G(s)~.
\end{equation}
Here, the potential $V(s)$ can be obtained from the SU(3) heavy-light chiral effective Lagrangian, which up to
next-to-leading order (NLO) contains 6 low-energy constants (LECs), as discussed below. Further,
$G(s)$ is the 2-point scalar loop function, which is best written in a dispersion-theoretical representation and
regularized with a subtraction constant $a(\mu)$, see e.g.~\cite{Oller:2000fj}. Of course, $T$, $V$ and
$G$ are matrices in the channel space, but for clarity I have suppressed the channel indices.
Let us now discuss the NLO effective chiral Lagrangian for $D\phi$ coupled channel dynamics. Following
Ref.~\cite{Guo:2008gp}, it takes the form
\begin{eqnarray}
{\cal L}_{\rm eff} &=& {\cal L}^{(1)} + {\cal L}^{(2)}~, \nonumber\\
{\cal L}^{(1)} &=& {\cal D}_\mu D {\cal D}^\mu D^\dagger - M_D^2 DD^\dagger~,\nonumber
\end{eqnarray}
\begin{eqnarray}
{\cal L}^{(2)} &=& D\,\bigl[-{h_0}\langle\chi_+ \rangle
 -{h_1} \chi_+ \nonumber\\
 &+& {h_2}\langle u_\mu u^\mu\rangle -{h_3} u_\mu u^\mu\bigr] D^\dagger\nonumber\\
 &+& {\cal D}_\mu D \bigl[{h_4} \langle u^\mu u^\nu\rangle
               - {h_5}\{u^\mu,u^\nu\}\bigr]\,{\cal D}_\nu D^\dagger~,
\end{eqnarray}  
with $D = (D^0,D^+,D_s^+)$, ${\cal D}_\mu$ is the chiral covariant derivative,  $M_D$ the $D$-meson mass
(in the chiral limit) and the conventional chiral building blocks $u_\mu \sim \partial_\mu \phi~,
\chi_+  \sim {\cal M}$, etc.\ are used. The LECs can be determined as follows: 
$h_0$ can be fixed from the pion-mass dependence of the $D$ and $D_s$ masses
and $h_1 = 0.42$ is given by the $D_s$-$D$ splitting. Further, $h_{2,3,4,5}$ are
obtained from a fit to lattice data ($D\pi\to D\pi, D\bar{K}\to D\bar{K}, ...$), see Ref.~\cite{Liu:2012zya}.
In that paper, 5 ``simple'' channels that do not contain  disconnected diagrams where fitted and then UCHPT
was used to predict the mass of the charm-strange scalar meson (the pole in the $(S,I) = (1,0)$ channel)
at $2315_{-28}^{+18}\,$ MeV, consistent with the PDG value for the $D_{s0}^\star(2317)$. To make contact to
the results from~\cite{Moir}, we need the appropriate formulation of UCHPT on a finite volume (FV).
\begin{figure}[t]
\centering
\includegraphics[width=0.40\textwidth]{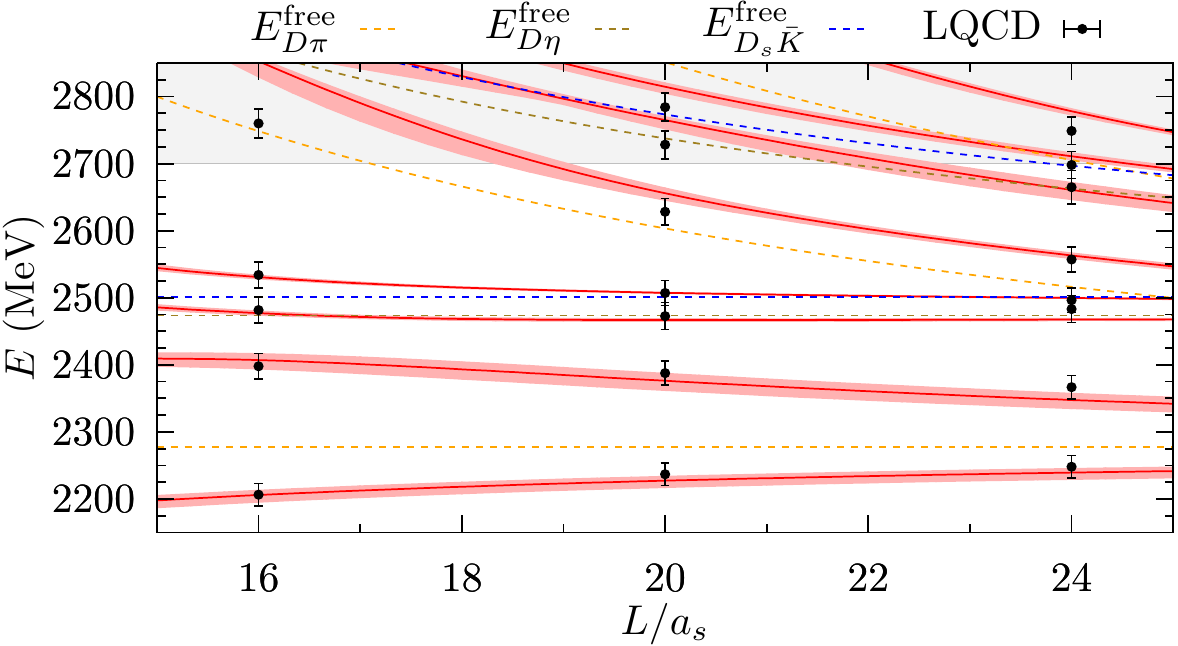}
\includegraphics[width=0.40\textwidth]{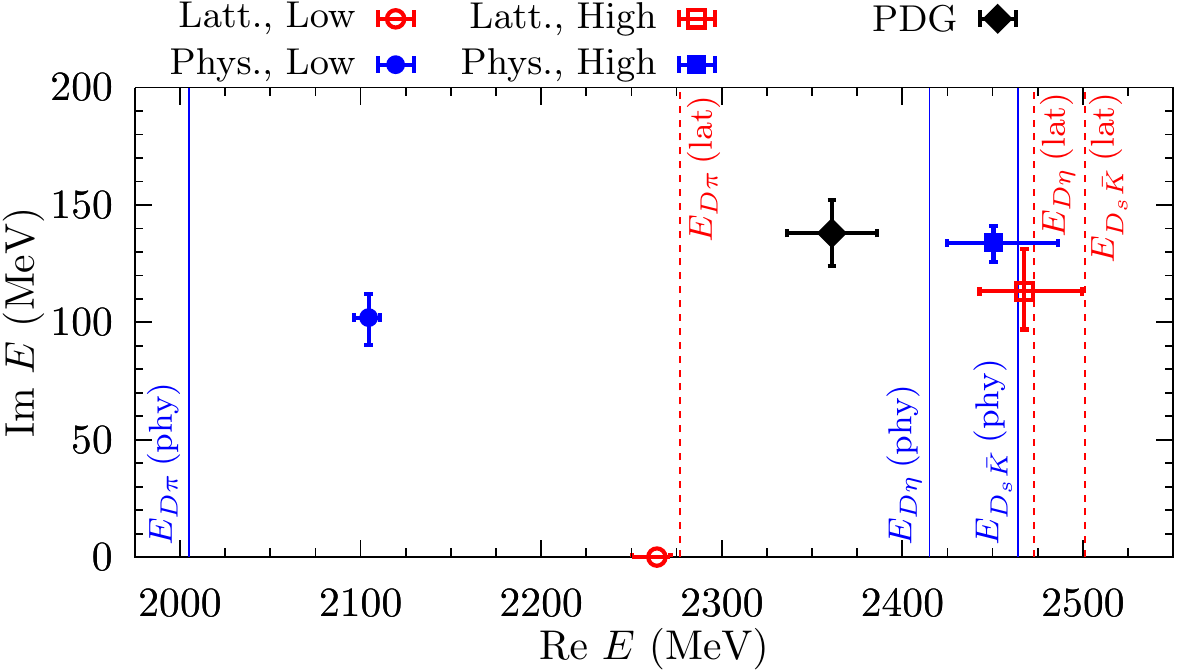}
\caption{The upper  panel shows the energy levels calculated in UCHPT with LECs determined in 2013
  in comparison to the results of~\cite{Moir} in the $(S,I)= (0,1/2)$ channel.
  Lower panel: Complex energy plane location of the two-pole-structure. Empty red (filled blue) symbols
  stand for the poles obtained when the LQCD~\cite{Moir}  (physical) masses are used. The black diamond
  represents the isospin average of the PDG values for $D_0^\star(2400)^0$ and $D_0^\star(2400)^+$. Figures
  from~\cite{Albaladejo:2016lbb}.}
\label{fig:Dtwo}
\vspace{-4mm}
\end{figure}
As noted before, in that case the momenta get quantized $\sim 1/L$ and the loop  function $G(s)$ gets modified.
Using the formalism of Ref.~\cite{Doring:2011vk}, we have (note the aforementioned regularization of
the L\"uscher function)
\begin{eqnarray}
\tilde{G}(s,L) = \displaystyle\lim_{\Lambda\to\infty}\left[ \frac{1}{L^3} \sum_{\vec n}^{|\vec{q}|<\Lambda}  I(\vec{q}\,) - \int_0^\Lambda \frac{q^2 dq}{2\pi^2} I(\vec{q}\,) \right]~,
\end{eqnarray}
where $I(\vec{q})$ is the integrand of $G(s)$.
The FV energy levels are obtained from the  poles of $\tilde{T} (s,L)$ via
$\tilde{T}^{-1}(s,L) = V^{-1}(s) - \tilde{G}(s,L)$.
Note that on a torus the potential $V(s)$ is the same as in the continuum. Using this framework, the
HSC data were re-analyzed in Ref.~\cite{Albaladejo:2016lbb}. First, the energy levels were postdicted
(no parameter needed to be fixed) to a high precision as shown in the upper panel of Fig.~\ref{fig:Dtwo}.
Second, and even more stunning, was the observation of the two-pole structure of the $D_0^\star (2400)$,
as depicted in the lower panel of Fig.~\ref{fig:Dtwo}, in complete analogy to the case of the
$\Lambda(1405)$~\cite{Oller:2000fj}. This solves the enigma that the mass of the $D_0^\star (2400)$,
which is made of a light ($u,d$) and a charm quark, is larger than the one of the $D_{s0}^\star(2317)$, which contains
the heavier strange quark. The lower of the two poles is lying visibly below the $D_{s0}^\star(2317)$!
This two-pole structure is easily understood from group theory. Consider the SU(3) limit, where all light
and heavy mesons take common values, analogous to the study of the $\Lambda(1405)$ in~\cite{Jido:2003cb}. 
Combining the $D$-meson anti-triplet with the Goldstone boson octet gives
\begin{equation}
  \bar{3}\otimes 8 = \underbrace{\bar{3} \oplus 6}_{\rm attractive} \oplus \overline{15}~,
\end{equation}
where the anti-triplet and the sextet are attractive, leading to two zero-width poles. Once the
SU(3) breaking in the meson masses is switched on, these poles move to the positions in the
complex energy plane shown in Fig.~\ref{fig:Dtwo}. We note that this two-pole structure had been
observed in earlier calculations~\cite{Kolomeitsev:2003ac,Guo:2006fu,Guo:2009ct,Guo:2015dha}
but did not receive the proper attention then. In fact, as pointed out in Ref.~\cite{Du:2017zvv},
there is an easy lattice test for this scenario: The sextet pole becomes a bound state 
for $M_\phi > 575\,$MeV in the SU(3) limit. Such a calculation could be easily done and would be
a fine test of the two-pole scenario. In fact, there is further phenomenological support of
this picture.
\begin{figure}[t]
\centering
\includegraphics[width=0.40\textwidth]{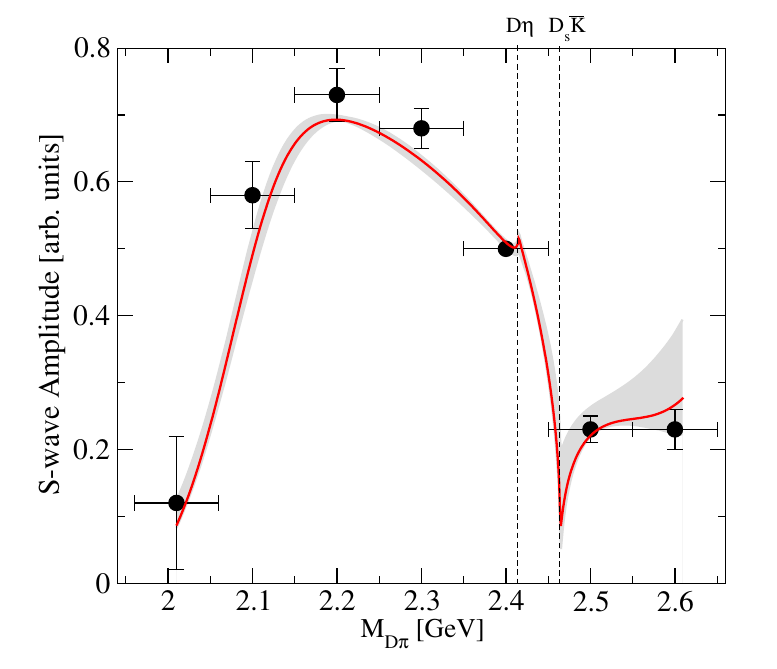}
\caption{Comparison of the S-wave amplitude determined from
the calculation of~\cite{Du:2017zvv} with the S-wave  anchor
points  from the  experimental analysis, shown as the data points~\cite{Aaij:2016fma}.
The red (solid) line gives  the  best  fit  results  and  the  grey  band  quantifies  the
uncertainties. The dashed perpendicular  lines indicate the location of the
$D\eta$ and $D_s \bar K$ threshold, respectively. Figure from~\cite{Du:2017zvv}.}
\label{fig:BDpipi}
\vspace{-4mm}
\end{figure}
Analyzing the high-precision data on the decays $B \to D\phi\phi$ from LHCb
\cite{Aaij:2015kqa,Aaij:2015vea,Aaij:2016fma}
within the same framework (using the pre-determined S-waves) leads to an excellent
description of the so-called angular moments~\cite{Du:2017zvv,Du:2019oki} and a prediction of two cusps
in the S-wave
at the $D\eta$ and $D_s \bar K$ thresholds, as shown in Fig.~\ref{fig:BDpipi}. Shown are the
so-called anchor points
provided by LHCb, where the strength and the phase of the S-wave were extracted from the data and
connected by a cubic spline in comparison to the UCHPT result of  Ref.~\cite{Du:2017zvv}. I point out that
the higher mass pole at 2.45~GeV  amplifies the predicted cusps.
In fact, the two-pole scenario can be extended to the axial-vector states and the corresponding $B$-mesons
using HQSS and HQFS, respectively. The emerging picture based on the calculations in Ref.~\cite{Du:2017zvv}
(and references therein) is summarized in Table~ \ref{tab:tp}.
Note that the PDG lists one state for the $D_0^\star$ and for the  $D_1$, these are located
at $(2318\pm 29, 134\pm 20)$~MeV and $(2427\pm 40, 192^{+65}_{-55})$~MeV, respectively. In view
of the results obtained in UCHPT, time is ripe to change these entries in the PDG. The predicted
two-pole scenarios for the $B_0^\star$ and the  $B_1$ are a nice test to further validate this
picture.

\begin{table}
\centering
\caption{Predictions of two states in various $I=1/2$ channels in the heavy meson sector. Here $(M, \Gamma/2)$
  denote the mass and the half-width, respectively, in units of [MeV]. From Ref.~\cite{Du:2017zvv}.}
  \label{tab:tp}       
\begin{tabular}{|l|cc|}
\hline
    & Lower pole  & Higher pole \\
\hline
  $D_0^\star$  & $\left(2105^{+6}_{-8}, 102^{+10}_{-11}\right)$  
             &  $\left(2451^{+36}_{-26}, 134^{+7}_{-8}\right)$ \\
 $D_1$       & $\left(2247^{+5}_{-6}, 107^{+11}_{-10}\right)$  
             & $\left(2555^{+47}_{-30}, 203^{+8}_{-9}\right)$ \\
 $B_0^\star$  & $\left(5535^{+9}_{-11}, 113^{+15}_{-17}\right)$  
             & $\left(5852^{+16}_{-19}, 36\pm 5\right)$ \\
 $B_1$       & $\left(5584^{+9}_{-11}, 119^{+14}_{-17}\right)$  
             & $\left(5912^{+15}_{-18}, 42^{+5}_{-4}\right)$ \\
\hline 
\end{tabular}
\end{table}

\section{Lesson 4: Hadronic molecules}
\label{sec-4}

As already stated in the introduction,  QCD offers yet another set of bound states, first seen in 
{\em nuclear physics}, namely {\bf hadronic molecules}, which are bound states made of 2 or 3 hadrons.
In what follows, I consider bound states of two hadrons in  S-wave very close to a 2-particle threshold
or  between two close-by thresholds. Such molecular states are weakly bound, i.e. the binding energies are
much smaller than the particle mass, and this weak binding also entails a large spatial extent.
Further, hadronic molecules show particular decay patterns. The classical example is, of course, the
deuteron, a bound state of a proton and a neutron. Its binding energy (BE) of about 2.22~MeV is much smaller
than its mass, $M_D= m_p + m_n - E_B \simeq 1876$~MeV and its radius of 2.14~fm is much bigger than
the proton radius of about 0.85~fm. Other examples are the two poles in the $\Lambda(1405)$ region, the
$f_0 (980)$, the $X(3872)$, and many others, for a recent review, see~\cite{Guo:2017jvc} and the discussion
below.

Naturally the question arises how to distinguish these molecules from compact multi-quark states?
This was originally answered by Weinberg~\cite{Weinberg:1965zz} and then refined by various others, see 
e.g.~\cite{Morgan:1990ct,Tornqvist:1994ji,Baru:2003qq}. For that, consider the wave function of a
bound state $|\Psi\rangle$  with a compact  component $|\psi_0\rangle$ and a two-hadron component 
$|h_1h_2\rangle$ in S-wave: 
\begin{equation}
|\Psi \rangle = \displaystyle\left(\sqrt{Z}|\psi_0\rangle\atop
\chi (\vec{k}) |h_1h_2 \rangle \right)~,
\end{equation}
that is, such a bound state consists of a compact component with  probability $\sqrt{Z}$ and a
two-hadron component with a relative wave function $\chi(\vec{k})$ (which should be normalizable).
Then, comparing the hadron-hadron scattering amplitude with the effective range expansion gives:
\begin{eqnarray}
\label{eq:comp}
a = -2 \frac{1-Z}{2-Z}\left(\frac{1}{\gamma} \right) + {\mathcal O}\left(\frac{1}{\beta} \right)~,\nonumber\\
r = -\frac{Z}{1-Z} \left(\frac{1}{\gamma} \right) + {\mathcal O}\left(\frac{1}{\beta} \right)~,
\end{eqnarray}
with $\gamma = \sqrt{2\mu E_B}$ the binding momentum, $\mu$ the reduced mass, $a$ the scattering 
length, $r$ the effective range and $\beta$ is the range of forces. For a pure molecule  ($Z=0$) we have 
the maximal scattering length, $a = -1/\gamma$, and  a natural effective range $r = {\mathcal O}(1/\beta)$.
Very differently for a compact state ($Z=1$), the scattering length  $a = - {\mathcal O}(1/\beta)$ is of natural
size and the effective range diverges, which is cured by higher order effects not explicitly
displayed in Eq.~(\ref{eq:comp}).
As an example, consider the deuteron. Here, we have $\gamma = 45.7\,$MeV$ ~= 0.23\,$fm$^{-1}$ and the
range of forces is set by the  one-pion-exchange, $1/\beta \sim 1/M_\pi \simeq 1.4\,$fm. Setting $Z=0$ in 
Eq.~(\ref{eq:comp}), we find $a_{\rm mol} = -(4.3\pm 1.4)\,$fm and $r_{\rm mol} = 1.4\,$fm, consistent 
with the data, $a = -5.419(7)\,$fm and $r = 1.764(8)\,$fm. 

\begin{figure}[t]
\centering
\includegraphics[width=0.220\textwidth]{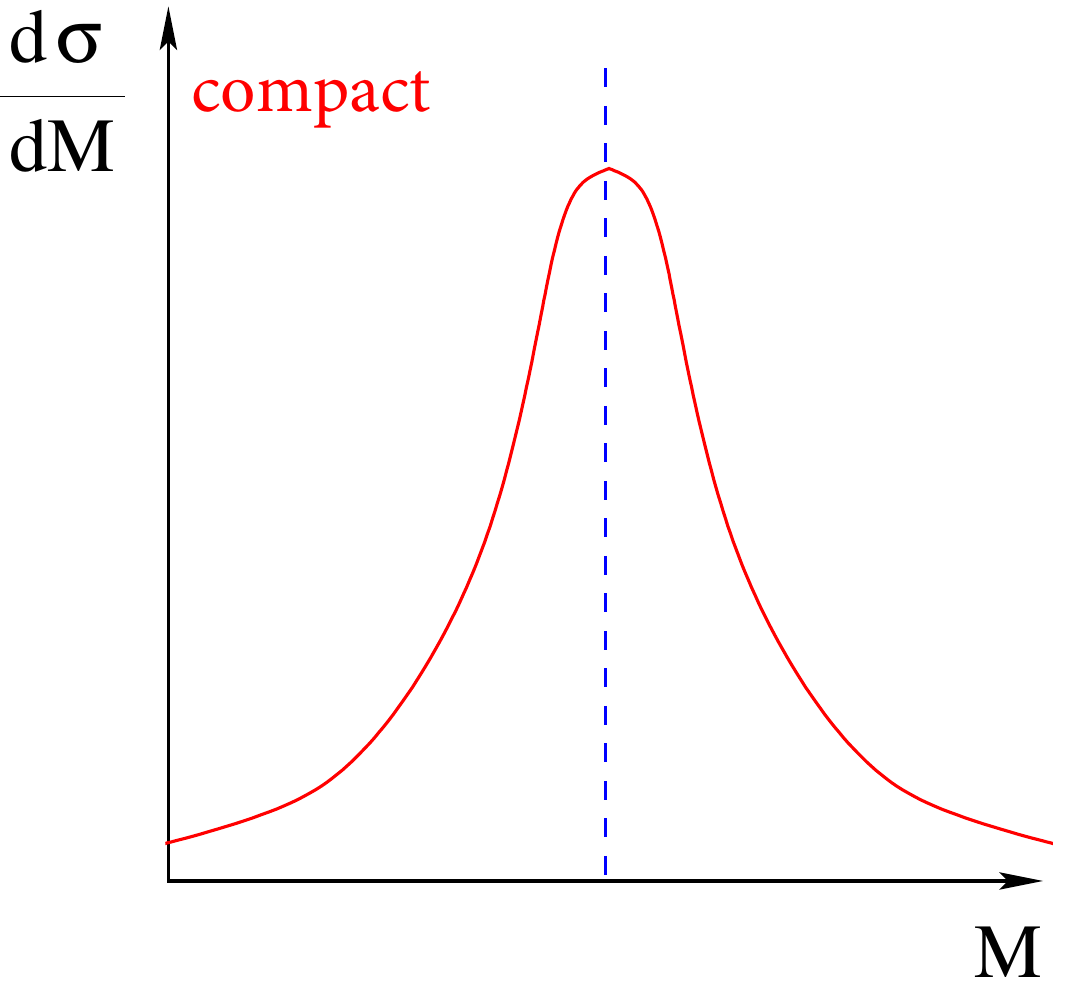}
\includegraphics[width=0.220\textwidth]{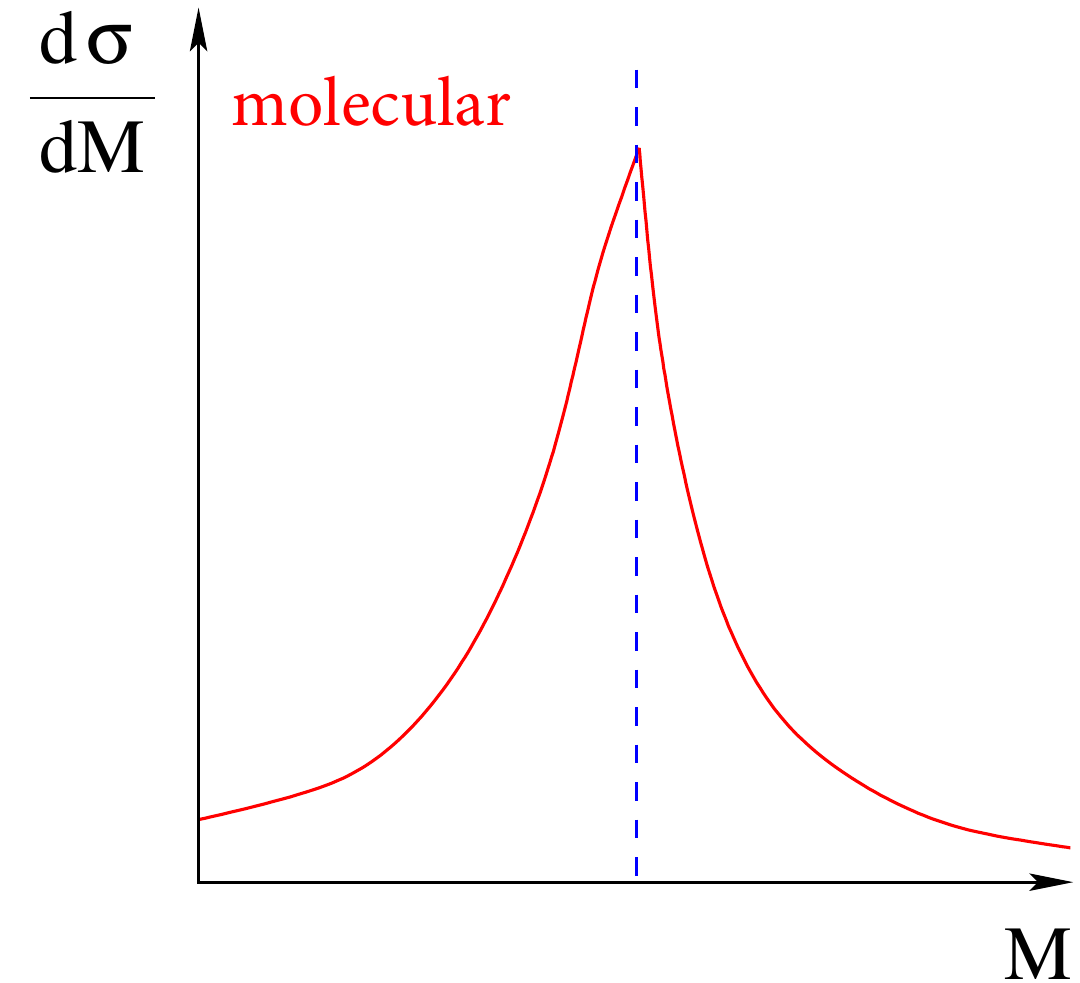}
\caption{Typical near-threshold line shapes that emerge for compact (left panel) and molecular states (right panel).
  The dashed perpendicular lines indicate the location of the threshold.  Note the cusp at the threshold for the
  molecular scenario. The $x$-axis shows 
  $M = m_1 + m_2 +E$. Figure inspired from~\cite{Guo:2017jvc}.}
\label{fig:lines}
\vspace{-4mm}
\end{figure}

Besides the deuteron all other (candidates for) hadronic molecules are unstable. Then the scattering T-matrix 
needs to be modified compared to the form discussed before. The corresponding modifications of the framework to
cope  with hadron resonances have been developed, 
see e.g.~Refs.~\cite{Baru:2003qq,Braaten:2007dw,Aceti:2012dd,Guo:2015daa}.
Assuming again closeness to a two-particle threshold, the scattering amplitude takes the form
\begin{equation}
{\rm T}(E) =  \displaystyle\frac{g^2/2}{E-E_R+(g^2/2)(i k+\gamma)+i\Gamma_0/2}~,
\end{equation}
with $g$ the coupling constant of the two-hadron system,  $E=k^2/(2\mu)$, $E_R$ the resonance energy 
and $\Gamma_0$  accounts for the inelasticities of other channels.  This leads to very different {\em  line shapes} 
for compact and molecular states. While for a compact state the  $k^2$ term dominates leading to a  symmetric
line shape, for a molecular state the  $g^2$ term dominates, leading to an asymmetric distribution exhibiting 
a cusp, cf. Fig.~\ref{fig:lines}. Further generalizations to instable constituents, virtual poles, etc., have also 
been worked out. I refer again to the review~\cite{Guo:2017jvc} for details.

Without going into details, let me list a few candidates for hadronic molecules (for details, see e.g. Ref.~\cite{Guo:2017jvc}).
Prominent examples in the light quark sector are the scalar mesons $f_0(980)$, $a_0(980)$ as well as the already
mentioned  two $\Lambda (1405)$ states, here I refer to the talks by Mai~\cite{MM} and Oset~\cite{EO} for further
discussion. There are also
very prominent examples in the $c\bar c$ and $b\bar b$ spectrum, these are the $X(3872)$, $Z_c(3900)$, $Y(4260)$,
$Y(4660)$,  $Z_b (10610)$, $Z_b (10650)$, and others. Similarly, in heavy-light systems we have a number of 
candidates, especially the $D_{s0}^\star (2317), D_{s1}(2460)$, $D_{s1}^\star(2860)$, and others. The molecular
nature of the charm-strange mesons was already discussed in Sect.~\ref{sec-3}.  In the baryon sector,
the LHCb pentaquarks \cite{Aaij:2015tga,Aaij:2019vzc} are good candidates, especially the recently
observed $P_c(4457)$, see e.g.
Refs.~\cite{Wu:2010jy,Chen:2015loa,Burns:2015dwa,Roca:2015dva,Chen:2019asm,Guo:2019fdo,Burns:2019iih}.

\begin{figure}[t]
\centering
\includegraphics[width=0.40\textwidth]{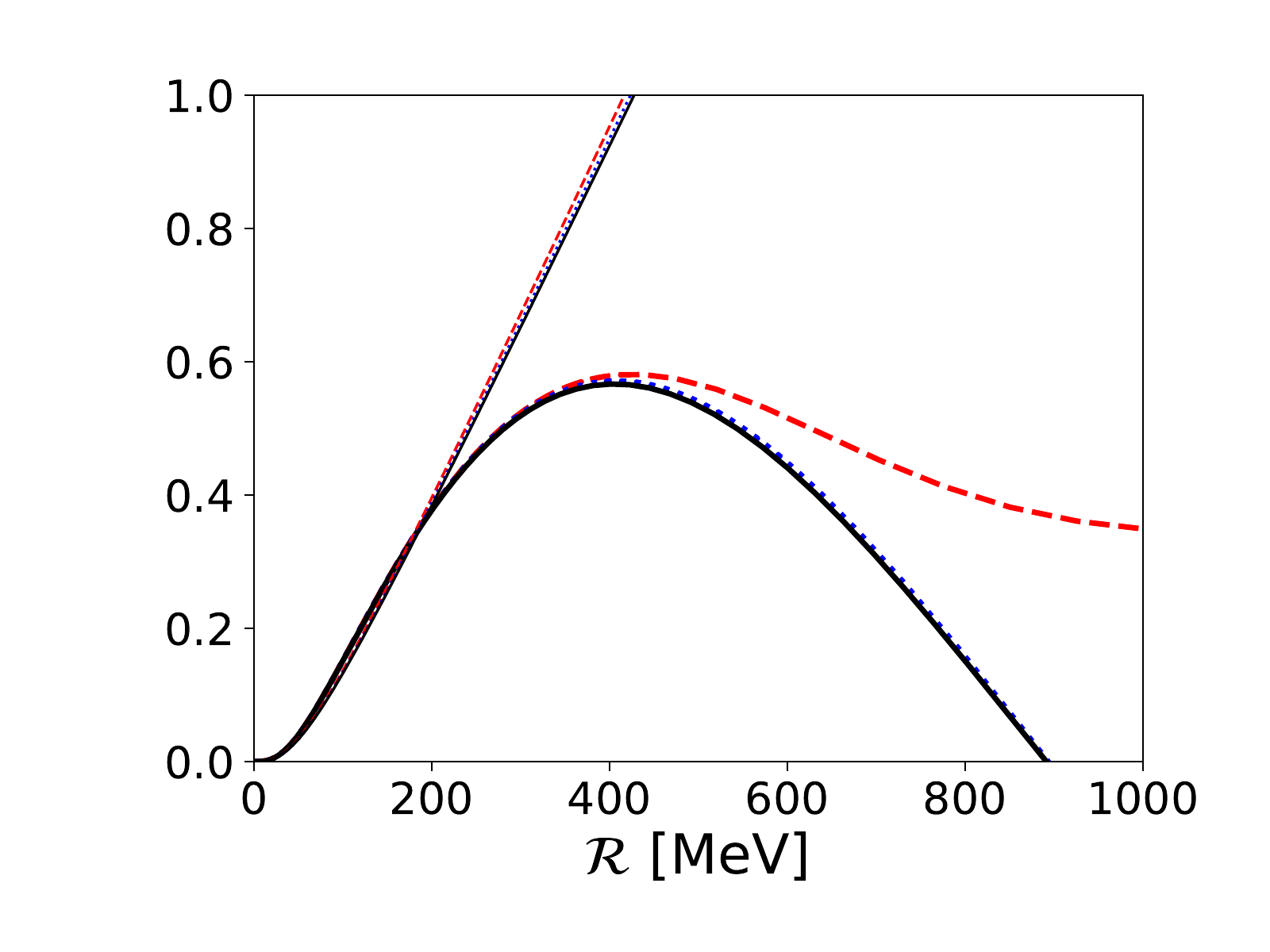}
\caption{Estimate of $\bar \Psi_\lambda({\cal R})$
for various deuteron wave functions: Results for $\lambda = 0.8, 1.6, 
4$~GeV are shown as red (dashed), blue (dotted) and black (solid) curves, respectively. 
The thick (thin) lines depict the results with (without) OPE. Figure from~\cite{Albaladejo:2017blx}.}
\label{fig:R}
\vspace{-4mm}
\end{figure}
I will take the opportunity to discuss a few misconceptions surrounding hadronic molecules. It is often claimed that 
molecules due to their large spatial extent can not  be produced in high-energy collisions, say at the LHC, but this
is not correct.  An argument found in the literature is due to Ref.~\cite{Bignamini:2009sk}. They consider the $X(3872)$
production at the Tevatron, and derived a bound on the cross section:
\begin{eqnarray}
\hspace{-0.85cm}\sigma(\bar pp\to X) &\sim&\!\!\! \left| \int d^3{\mathbf k}\langle X|D^0\bar D^{*0}
({\mathbf k})\rangle\langle D^0\bar D^{*0}({\mathbf k})|\bar pp\rangle\right|^2
\nonumber \\
&\simeq&\!\!\!  \left| \int_{\cal R} d^3{\mathbf k}\langle X|D^0\bar D^{*0}({\mathbf
k}) \rangle\langle D^0\bar D^{*0}({\mathbf k})|\bar pp\rangle\right|^2 \nonumber
\\
&\leq& \!\!\! \int_{\cal R} d^3 {\mathbf k} \left|\Psi({\mathbf k})\right|^2
\int_{\cal R} d^3 {\mathbf k}\left|\langle D^0\bar D^{*0}({\mathbf k})|\bar
pp\rangle\right|^2 \nonumber \\
 &\leq&\!\!\!  \int_{\cal R} d^3 {\mathbf k}\left|\langle D^0\bar D^{*0}({\mathbf
k})|\bar pp\rangle\right|^2  \, ,
\label{eq:bi}
\end{eqnarray}
which depends crucially on the value of $\cal R$ which specifies the region where the bound state wave 
function ``$\Psi(\mathbf k)$  is significantly different from zero''~\cite{Bignamini:2009sk}. These authors
assume that  ${\cal R}$ is of the order of the binding momentum $\gamma$, which amounts to  ${\cal R}\simeq 35$~MeV 
for the then accepted BE of the $X(3872)$. From this they conclude that $\sigma(\bar pp\to X) \simeq 0.07$~nb, which is orders
of magnitudes lower  than the  experimental bounds known then (updated numbers will be given below). So this
leads them to conclude that the  $X(3872)$ can not be a molecule. But I said that the  $X(3872)$  is a premier
candidate for a molecule in the charmonium spectrum, so what goes wrong with this argument? Let us take a step
back and consider the best known molecule, the deuteron~\cite{Albaladejo:2017blx}. The relevant integral is
\begin{equation}
\bar \Psi_\lambda({\cal R}) \equiv \int_{\cal R} d^3 {\mathbf k} \, \Psi_\lambda({\mathbf k})~,
\end{equation}
where $\lambda$ specifies a regulator that needs to be introduced to render the wave function well defined.
In the deuteron case, $\gamma \simeq 45\,$MeV, and as in the case of the $X(3872)$ is of the order of the
pion Compton wavelength ${\cal O}(1/M_\pi)$. In Fig.~\ref{fig:R}, $\bar \Psi_\lambda({\cal R})$  is shown for the deuteron 
with two sets of wave functions, one generated from a potential with a short-ranged term and one-pion exchange 
(OPE) and the other without OPE. For the former a Gaussian regulator is used and the small 
$D$-wave component is not shown. For the latter a sharp momentum space
cut off is used for simplicity. One sees that $\bar \Psi_\lambda({\cal R})$ is far
from being saturated for ${\cal R}\simeq\gamma_d$ for all values of 
$\lambda$. A much larger value of 
${\cal R}\sim$300~MeV$\sim 2M_\pi$ needs to be taken for the second line in
Eq.~(\ref{eq:bi}) to be a good approximation of the first.  This is, of course, related to the
fact that two-pion exchange is required to bind the deuteron.
The same  should also be true for the $X(3872)$ as a $D\bar D^*$ 
molecule, since the range of forces is the same. A similar value was also favored in
Ref.~\cite{Artoisenet:2009wk} based on rescattering arguments. In fact, 
Ref.~\cite{Bignamini:2009sk} noted that with such a value of $\cal R$ the 
upper bound becomes consistent with the CDF result. 

\begin{figure}[t]
\centering
\includegraphics[width=0.30\textwidth]{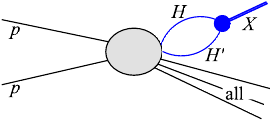}
\caption{The mechanism for the inclusive production of the $X$ as a $HH^{\prime}$ 
bound state in proton--proton collisions. Here, $all$
denotes all the produced particles other than the $H$ and $H^{\prime}$ in the
collision. Figure from~\cite{Guo:2014sca}.}
\label{fig:prod}
\vspace{-4mm}
\end{figure}
Let me briefly discuss how  one can calculate the hadroproduction of the $X(3872)$ in an EFT if one realizes that this
is a process involving short-distance physics, but still  factorization is at work~\cite{Artoisenet:2009wk}. 
In such a scheme, event generators are used to calculate the collinear production of two hadrons and
these then combine to form a molecule at large distances, see Fig.~\ref{fig:prod}, which means that the molecule is
not promptly produced. 
For high-energy processes like at the Tevatron or the LHC, the corresponding cross section of producing
a molecule $X$ from two hadrons $H,H^\prime$ can be calculated
via~\cite{Guo:2013ufa,Guo:2014sca}:
\begin{equation}
\sigma[X] =  \displaystyle\frac{1}{4m_Hm_{H'}} g^2 |G|^2
\bigg(\frac{d\sigma[HH^{\prime}(k)]}{dk}\bigg)_\text{MC}\frac{4\pi^{2}\mu}{k^{2}}~,
\nonumber\\[-3ex]
\end{equation}
\begin{eqnarray}
G(E,\Lambda) &=&  -\displaystyle\frac{\mu}{\pi^2}\bigg[\sqrt{2\pi}\,\frac{\Lambda}{4}+\sqrt{\pi}\,\gamma
	D\left(\frac{\sqrt{2}\gamma}{\Lambda}\right)\nonumber\\
       & &-\frac{\pi}{2}\,\gamma\,e^{2\gamma^2/\Lambda^2}\bigg]~,
\end{eqnarray}
with $(d\sigma[HH^{\prime}(k)]/dk)_\text{MC}$ the cross section of producing the hadrons $H$ 
and $H'$ in collinear geometry, $G$ is the cut-off regularized two-meson ($HH'$) loop function 
expressed in terms of the Dawson function, $D(x)=e^{x^2}\,\int^{x}_0\,e^{-y^2}\,dy$, 
$\gamma=\sqrt{-2\mu(E-m_H-m_{H'})}$ is the binding momentum and $\Lambda$ is the
cutoff. Following Ref.~\cite{Guo:2013sya}, a range of $[0.5,1.0]$~GeV is used for the cutoff $\Lambda$.
Typical results using the PYTHIA and HERWIG generators are collected in Tab.~\ref{tab:Xp}.
\begin{table}[t]
\caption{Integrated cross sections (in units of nb) for   $pp/\bar p\to X(3872)$
compared to  experimental measurements by CDF~\cite{Bauer:2004bc}
and CMS~\cite{Chatrchyan:2013cld}.}
\begin{tabular}{|c|cc|}
\hline
$\sigma(pp/\bar{p}) \!\!\to\!\! X(3872))$  & $\Lambda \in [0.5 ,1.0]$\,GeV &\!\!\!\!\!\!\!\! Exp.\\
\hline
Tevatron &  \!\!\!\!\!\! 5 - 29 [nb]& \!\!\!\!\!\!\!\!37 - 115 [nb]\\
LHC7     &  \!\!\!\!\!\! 4 - 55 [nb]& \!\!\!\!\!\!\!\! 13 - 39 [nb] \\
\hline
\end{tabular}
\label{tab:Xp}
\end{table}
We see that the calculations are not very precise, but they are perfectly consistent with the data,
very different from the conclusions drawn in Ref.~\cite{Bignamini:2009sk}. For a similar calculation
concerning the hadroproduction of the charm-strange mesons, I refer to Ref.~\cite{Guo:2014ppa}.

\begin{figure}[t]
\centering
\includegraphics[width=0.30\textwidth]{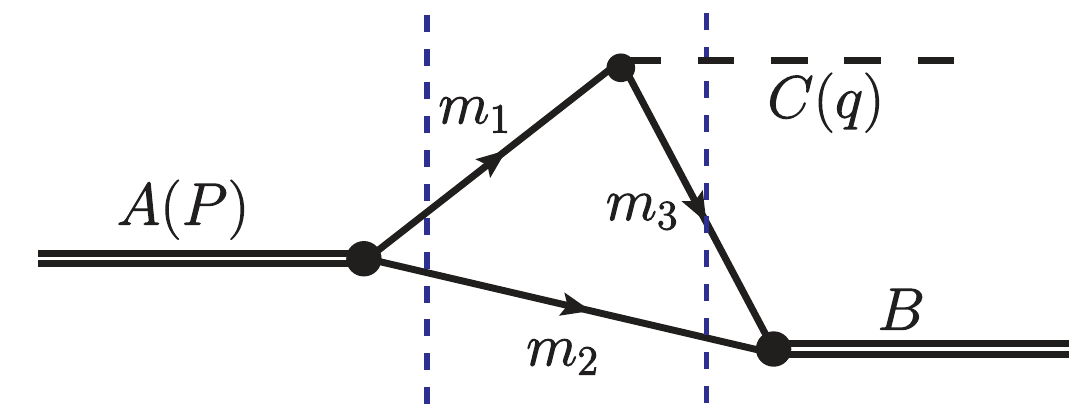}
\caption{A triangle diagram illustrating the long-distance contribution to the transition between 
two heavy particles A and B with the emission of a light particle C. The two vertical dashed lines 
denote the two relevant cuts. Figure from~\cite{Guo:2017jvc}.}
\label{fig:tri}
\vspace{-4mm}
\end{figure}
Finally, I mention that most  candidates  for hadronic molecules have been found through decays that often
involve triangle diagrams with an anomalous singularity, as studied by Landau, Nambu and
others in the 1950ties.  There exist  different EFTs that are used in the calculations of hadronic
molecules and their decays. These are the already mentioned UCHPT as well as NREFT$_1$ and NREFT$_2$ 
when at least one heavy quark is involved. In NREFT$_1$, all intermediate particles in the triangle diagram in
Fig.~\ref{fig:tri} are close to their mass shell, so that one can expand in  powers of the average velocity and 
external (small) momenta. This has e.g. been applied to systematic studies of a number of charmonium transitions.
In  NREFT$_2$ one intermediate particle is further off its mass shell, so one integrates out this particle and then 
proceeds as before. This EFT was originally invented as XEFT for studies of the $X(3872)$~\cite{Fleming:2007rp}.
XEFT resembles much the pionless EFT of nuclear physics, see e.g.~\cite{Bedaque:2002mn}.
It has been used for systematic studies of processes involving the $X(3872)$ and the $Z_b$ states,
see e.g.~\cite{Canham:2009zq}.

\section{The width of the lightest nucleon resonances from baryon CHPT}
\label{sec-5}

In this section, I will be concerned with calculating the width of the two lightest
baryon resonances, the $\Delta(1232)$ and the Roper $N^*(1440)$. This might
at first sight appear irritating, as imaginary parts are usually not precisely
reproduced in CHPT. For that simple reason, one has to employ a complex-mass
scheme and work to two loops. The complex-mass renormalization scheme is  a method that was
originally introduced for precision $W,Z$-physics, see e.g.~\cite{RGS,Denner:1999gp}  and
later transported to chiral EFT~\cite{Djukanovic:2009zn}.

Let me give a brief outline of the complex-mass scheme (CMS), following Ref.~\cite{Denner:2006ic}.
Consider first an instable particle at tree level. The CMS amounts to treating the mass of this particle
consistently  as a complex quantity, defined as the location of the pole in
the complex $k^2$-plane of the corresponding propagator with momentum $k$. It can be shown that
this scheme is symmetry-preserving and leaves the corresponding Ward identities intact.  Extending 
this to one loop, one splits the real bare masses into complex renormalized masses and complex counter\-terms.
This is important, as only renormalized masses are observable. The corresponding Lagrangian yields Feynman
rules with complex masses and counterterms, which allows for standard perturbative calculations. This is
essentially a rearrangement of contributions that is not affected by double counting. The imaginary part of the
particle mass appears in the propagator and is resummed in the Dyson series. In contrast to this, the
imaginary part of the counterterm is not resummed. One can show that in such a case gauge invariance remains
valid, and unitarity cancellations are respected order by order in the perturbative expansion. This also requires 
integrals with complex internal masses, as worked out in Ref.~\cite{Beenakker:1988jr}. For further discussions\
of the method, the reader is referred to Ref.~\cite{Denner:2006ic} (and references therein). In case of a
chiral EFT, the perturbative expansion proceeds as usual in terms of small momenta and quark masses, 
with a proper treatment of the heavy particle mass in loop diagrams (like the heavy-baryon scheme
or the so-called infrared-regularization or the extended-on-mass-scheme discussed below).

\subsection{The width of the {\boldmath$\Delta(1232)$}}
\begin{figure}[t]
\centering
\includegraphics[width=0.49\textwidth]{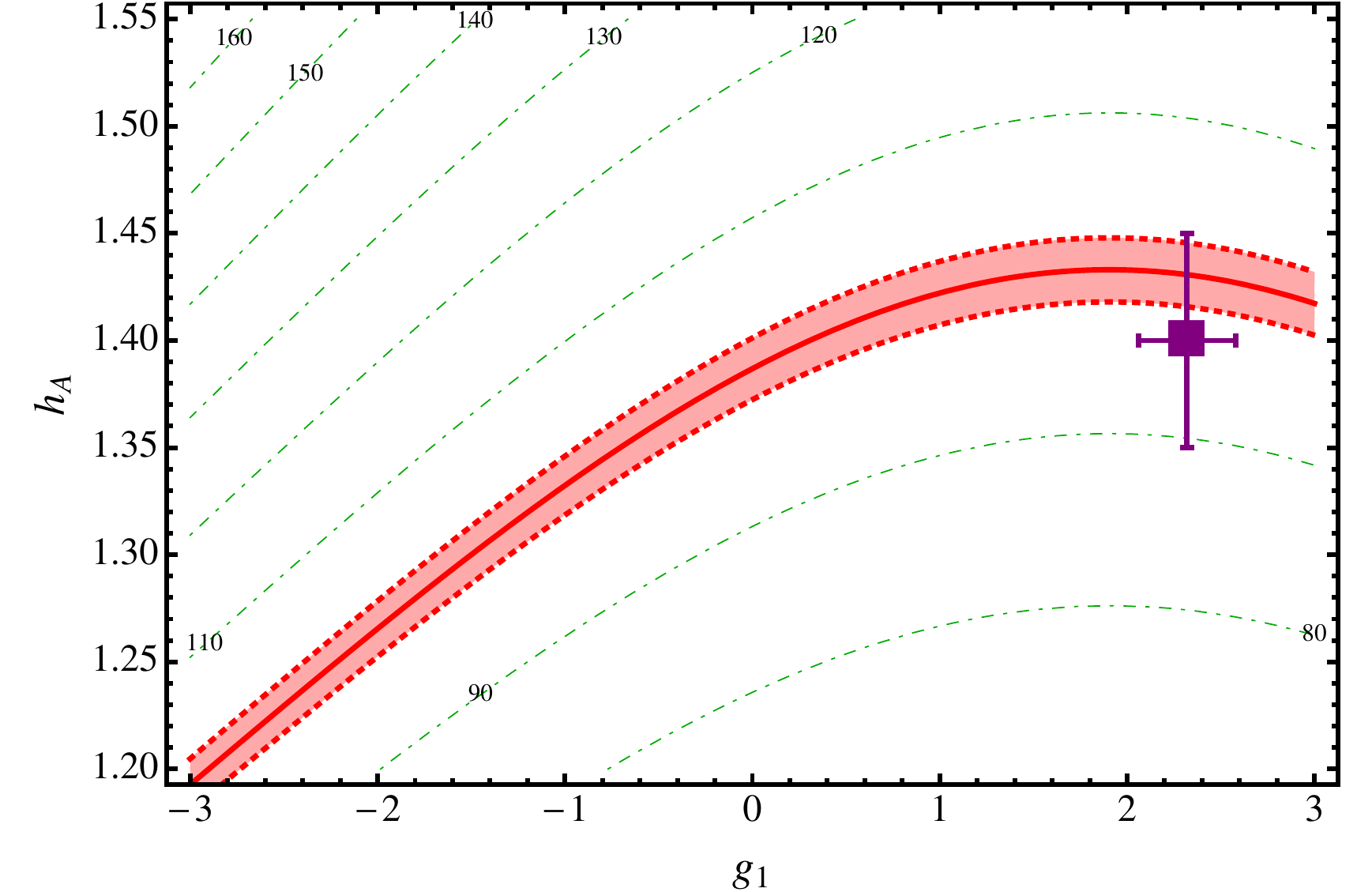}
\caption{Correlation between the leading $\pi \Delta$ and $\pi N \Delta$ couplings.
The central line corresponds to $\Gamma_\Delta = 100$~MeV  while the band is obtained by varying 
$\Gamma_\Delta$ in the range of 98 to 102~MeV. The dot-dashed lines show the correlation
for other values of the width of the Delta.
The box with the error bars are the results from the analysis of Ref.~\cite{Yao:2016vbz}.
Figure inspired from~\cite{Gegelia:2016pjm}.}
\label{fig:corr}
\vspace{-4mm}
\end{figure}

Consider first the width of the $\Delta$ at two-loop order~\cite{Gegelia:2016pjm}.
The  pertinent effective Lagrangian contains, besides many
other terms, the leading $\pi \Delta$ and $\pi N\Delta$ couplings,
parametrized in terms of the LECs $g_1$ and $h$, respectively,
\begin{eqnarray}
{\cal L}^{(1)}_{\pi\Delta} &=& -\bar{\Psi}_{\mu}^i\xi^{\frac{3}{2}}_{ij}\Bigl\{\left(i\slashed{D}^{jk}
-m_{\Delta}\delta^{jk}\right)g^{\mu\nu}\nonumber\\
&-&i\left(\gamma^\mu D^{\nu,jk}+\gamma^\nu D^{\mu,jk}\right) +i \gamma^\mu\slashed{D}^{jk}\gamma^\nu\nonumber\\
&+&m_{\Delta}\delta^{jk} \gamma^{\mu}\gamma^\nu
+ {g_1}\frac{1}{2}\slashed{u}^{jk}\gamma_5g^{\mu\nu}\nonumber\\
&+&g_2\frac{1}{2} (\gamma^\mu u^{\nu,jk}
+u^{\nu,jk}\gamma^\mu)\gamma_5\nonumber\\
&+& g_3\frac{1}{2}\gamma^\mu\slashed{u}^{jk}\gamma_5\gamma^\nu \Bigr\}\xi^{\frac{3}{2}}_{kl}
{\Psi}_\nu^l~,\nonumber\\
{\cal L}^{(1)}_{\pi N\Delta} &=& h\,\bar{\Psi}_{\mu}^i\xi_{ij}^{\frac{3}{2}}
\Theta^{\mu\alpha}(z_1)\ \omega_{\alpha}^j\Psi_N+ {\rm h.c.}~,\nonumber\\
{\cal L}^{(2)}_{\pi N\Delta}&=&\bar{\Psi}_{\mu}^i\xi_{ij}^{\frac{3}{2}}\Theta^{\mu\alpha}(z_2)\nonumber\\
&\times&\left[i\,b_3\omega_{\alpha\beta}^j\gamma^\beta+i\,\frac{b_8}{m}\omega_{\alpha\beta}^ji\,D^\beta\right]
\Psi_N+{\rm h.c.}\ ,\nonumber
\end{eqnarray}
\begin{eqnarray}
{\cal L}^{(3)}_{\pi N\Delta}&=&\bar{\Psi}_{\mu}^i\xi_{ij}^{\frac{3}{2}}\Theta^{\mu\nu}(z_3)\biggl[
  \frac{f_1}{m}[D_\nu,\omega_{\alpha\beta}^j]\gamma^\alpha i\,D^\beta\nonumber\\
&-&\frac{f_2}{2m^2}[D_\nu,\omega_{\alpha\beta}^j]
\{D^\alpha,D^\beta\}\nonumber\\
&+&f_4\omega_\nu^j\langle\chi_+\rangle+f_5[D_\nu,i\chi_-^j]\biggr]\Psi_N+ {\rm h.c.},
\label{eq:delta}
\end{eqnarray}  
where $\Psi_N$ and $\Psi_\nu$ are the isospin doublet field of the nucleon  
and the vector-spinor isovector-isospinor
Rarita-Schwinger field  of  the $\Delta$-resonance
with bare masses $m$ and $m_{\Delta 0}$, respectively. 
$\xi^{\frac{3}{2}}$ is the isospin-$3/2$ projector, 
$\omega_\alpha^i=\frac{1}{2}\,\langle\tau^i u_\alpha \rangle$ and $\Theta^{\mu\alpha}(z)=g^{\mu\alpha}
+z\gamma^\mu\gamma^\nu$. Using field redefinitions the off-shell parameters $z$  can be absorbed in 
LECs of other terms of the effective Lagrangian and therefore they can be chosen arbitrarily 
\cite{Tang:1996sq,Krebs:2009bf}. We fix the off-shell structure 
of the interactions with the delta by adopting $g_2=g_3=0$ and $z_1=z_2=z_3=0$.
For vanishing external sources, the covariant derivatives are given by 
\begin{eqnarray}
D_\mu \Psi_N & = & \left( \partial_\mu + \Gamma_\mu 
\right) \Psi_{N}\,,  \nonumber\\
\Gamma_\mu  &=& 
\frac{1}{2}\,\left[u^\dagger \partial_\mu u +u
\partial_\mu u^\dagger 
\right]=\tau_k\Gamma_{\mu,k}, \nonumber\\
\left(D_\mu\Psi\right)_{\nu,i} & = &
\partial_\mu\Psi_{\nu,i}-2\,i\,\epsilon_{ijk}\Gamma_{\mu,k} \Psi_{\nu,j}+\Gamma_\mu\Psi_{\nu,i}
\,. \label{cders}
\end{eqnarray}
\begin{figure}[t]
\centering
\includegraphics[width=0.50\textwidth]{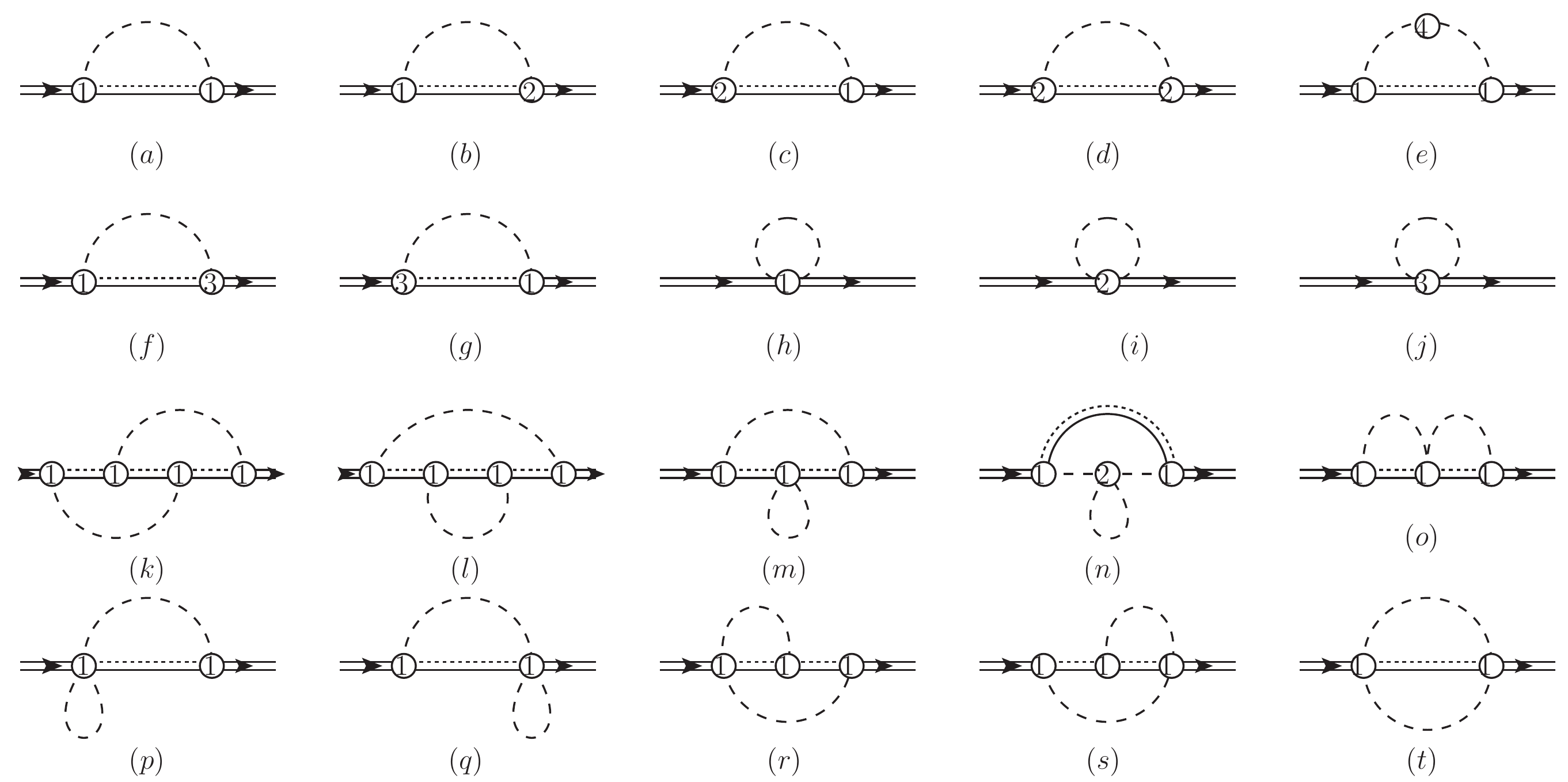}~~~~~
\caption{One and two-loop self-energy diagrams contributing to the width 
of the delta resonance up-to-and-including 
fifth order according to the standard power counting. The dashed and double solid lines 
represent the pions and the delta resonances, respectively. 
The double (solid-dotted) lines in the loops correspond to either nucleons or deltas. 
The numbers in the circles give the chiral orders of the vertices. Figure from~\cite{Gegelia:2016pjm}.}
\label{fig:Delta}
\vspace{-4mm}
\end{figure}
The power counting  rests on $m_\Delta - m_N$ being a small quantity.
More precisely, the small parameters are the external momenta, the pion mass
and the nucleon-Delta mass splitting, collectively denoted as $q$.
However, there are so many LECs in Eq.~(\ref{eq:delta}), so how can one
one possibly make a prediction?  Let us
evaluate the $\Delta$ self-energy on the complex pole,
\begin{equation}
  z - m_{\Delta}^0 - \Sigma(z) =0 \quad {\rm with} \quad
  z=m_\Delta-i\,\displaystyle\frac{\Gamma_\Delta}{2}~.
\end{equation}
The corresponding diagrams for the one- and two-loop self-energy contributing to the 
width of the delta resonance up to order $q^5$ are  displayed in Fig.~\ref{fig:Delta},
where the counter\-term diagrams are not shown. The one-loop diagrams are easily worked out.
For the calculation of the two-loop graphs one uses the Cutkosky rules for instable
particles, that relate the width to the pion-nucleon scattering amplitude,
$\Gamma_\Delta \sim |A(\Delta\to N\pi)|^2$~\cite{Veltman:1963th}.
One finds a remarkable  reduction of parameters that is reflected in the relation
\begin{eqnarray}
h_A &=& h - \left(b_3\Delta_{23}+b_8\,\Delta_{123}\right)\nonumber\\
&-&\left(f_1\Delta_{23}+ f_2\,\Delta_{123}\right)\Delta_{123}
+2(2f_4-f_5)M_\pi^2~,\nonumber\\
\Delta_{23} &=&m_N-m_\Delta~,\nonumber\\
\Delta_{123}&=& \frac{M_\pi^2+m_N^2-m_\Delta^2}{2m_N}~,
\end{eqnarray}  
which means that all of the LECs appearing in the $\pi N\Delta$ interaction at second
and third order, the $b_i \,(i=3,8)$ and $f_i \, (i=1,2,4,5)$, respectively, merely lead to a renormalization
of the LO $\pi N\Delta$ coupling $h$,
and, consequently, one finds a very simple formula for the decay width $\Delta \to N\pi$,
\begin{equation}
\Gamma(\Delta\to N \pi) =  (53.9\,{h}_A^2+0.9g_1^2 {h}_A^2-3.3g_1^{} {h}_A^2 
-1.0\,{h}_A^4)~{\rm MeV}~.
\end{equation}
This leads to a novel correlation that is independent of the number of colors, as $N_c$ was not
used as a parameter in the calculation. This correlation between $h_A$ and $g_1$ is depicted in
Fig.~\ref{fig:corr}. It is obviously fulfilled by the analysis of  Ref.~\cite{Siemens:2016jwj}, that showed that 
the inclusion of the $\Delta$ alleviates the tension between the threshold and subthreshold regions in the
description of $\pi N$ scattering found in baryon CHPT.

\subsection{The width of the Roper resonance}
Next, I consider the calculation of the width of the Roper $N^*(1440)$ at two-loop order~\cite{Gegelia:2016xcw}.
A remarkable feature of the Roper is the fact that its decay width into a nucleon and a pion is similar to the width
into a nucleon and two pions. Any model that is supposed to describe the Roper must account for this fact. In CHPT,
consider the  effective chiral Lagrangian of pions, nucleons and
deltas coupled to the Roper~\cite{Borasoy:2006fk,Djukanovic:2009gt,Long:2011rt},
\begin{eqnarray}
{\cal L}_{\rm eff}&=&{\cal L}_{\pi\pi}+{\cal L}_{\pi N}+{\cal L}_{\pi \Delta}+{\cal L}_{\pi R}\nonumber\\
&+ &{\cal L}_{\pi N\Delta}+{\cal L}_{\pi NR}+{\cal L}_{\pi\Delta R}~,
\end{eqnarray}
with
\begin{eqnarray}
{\cal L}_{\pi R}^{(1)}&=&\bar{\Psi}_R\left\{i\slashed{D}-m_R+\frac{1}{2}{g_R}\slashed{u}
\gamma^5\right\}\Psi_R~,\nonumber\\
{\cal L}_{\pi R}^{(2)} &=& \bar{\Psi}_R\left\{c_1^R\langle\chi_+\rangle\right\}\Psi_R + \ldots~,\nonumber
\end{eqnarray}
\begin{eqnarray}
  {\cal L}_{\pi NR}^{(1)} &=& \bar{\Psi}_R\left\{\frac{1}{2}{g_{\pi NR}}\gamma^\mu\gamma_5 u_\mu\right\}\Psi_N
+ {\rm h.c.}~, \nonumber\\
{\cal L}^{(1)}_{\pi \Delta R} &=& {h_R}\,\bar{\Psi}_{\mu}^i\xi_{ij}^{\frac{3}{2}}\Theta^{\mu\alpha}(\tilde{z})\ \omega_{\alpha}^j\Psi_R+ {\rm h.c.}~,
\end{eqnarray}
where  $g_R$, $g_{\pi NR}$ and $h_R$, respectively, are the leading Roper-pion, Roper-nucleon-pion
and Delta-Roper-pion couplings. Here, ${\Psi}_R$ denotes the Roper isospin doublet field and all
other notations are as in the preceeding subsection and in~\cite{Gegelia:2016xcw}.

In this case, the power counting is more complicated, but can be set
up around the complex pole as (for more details, see~\cite{Gegelia:2016xcw}), assigning
the following counting rules:
\begin{equation}
m_R-m_N \sim \varepsilon~,~~ m_R-m_\Delta \sim \varepsilon^2~,~~ m_\Delta-m_N \sim \varepsilon^2~,~~
M_\pi \sim \varepsilon^2~,
\end{equation}
where $\varepsilon$ denotes a small parameter. Again, let us calculate the self-energy to two loops
at the complex pole $z_R = m_R - i\Gamma_R/2$. By applying the cutting rules to these self-energy diagrams,
one obtains the graphs contributing to the decay amplitudes of the Roper resonance into the $\pi N$
and $\pi\pi N$ systems, leading to the total width
\begin{equation}
\Gamma_R = \Gamma_{R\to N\pi} +  \Gamma_{R\to N\pi\pi}~.
\end{equation}  
A somewhat lengthy calculation leads to:
\begin{equation}
\Gamma(R\to N \pi) = 550(58) \, g_{\pi NR}^2\ {\rm MeV}~,
\end{equation}
and
\begin{eqnarray}
& &\!\!\!\!\!\!\!\!\!\!\!\!\Gamma(R\to N\pi\pi) =\Bigl(1.5(0.6)\,g_A^2 \,g_{\pi NR}^2\nonumber\\
& & -2.8(1.0)\, g_A^{} \, g_{\pi NR}^2\,g_R^{}\\
& & +1.5(0.6)\,g_{\pi NR}^2\, g_R^2 + 3.0(1.0)\,g_A^{}\, g_{\pi NR}^{} \,h_A^{} h_R^{}\nonumber\\
& & -3.8(1.4)\,g_{\pi NR}^{}\,g_R^{} \,h_A^{} h_R^{} +9.9(5.5)\,h_A^2h_R^2\Bigr)~{\rm MeV}.\nonumber
\end{eqnarray}
The total width thus depends on five LECs. The uncertainties in the round brackets are generated by the
uncertainties in the LECs. We use $g_A =1.27$ and $h_A=1.42\pm 0.02$. The latter value is the real
part of this coupling taken  from Ref.~\cite{Yao:2016vbz}.
As for the other unknown parameters,  the authors of~\cite{Gegelia:2016xcw} fixed $g_{\pi NR}$ so as to
reproduce the width  $\Gamma_{R\to \pi N}=(123.5\pm 19.0)$~MeV from the PDG. 
This yields $g_{\pi NR}=\pm (0.47\pm 0.04)$. In what follows, let us take the
positive sign for our central value and use the negative one as part of the
error budget. Further,  assume $g_R=g_A$ and $h_{R}=h_A$, the so-called maximal mixing assumption~\cite{Beane:2002ud}.
Then, one can make a prediction for the two-pion decay width of the Roper,
\begin{equation}
\Gamma (R\to N\pi\pi) = (41 \pm 22_{\rm LECs} \pm 17_{\rm h.o.})~{\rm MeV}~,
\end{equation}  
which is consistent with the PDG value of  ($67\pm 10$)~MeV. The error due to the neglect of the higher orders (h.o.)
is simply given by multiplying the $\varepsilon^5$ result (central value) with $\varepsilon = (m_R-m_N)/m_N
\simeq 0.43$. Clearly, to make further progress, we need an improved determination of the LECs $g_R$ and $h_R$.
This could be addressed within LQCD. Finally, I would like to mention that this scheme has also
been used to consider the electromagnetic transition form factors of the Roper~\cite{Gelenava:2017mmk}.

\section{On the pion cloud of the nucleon and other hadrons}
\label{sec-6}

At this conference, one often encountered the situation that some observable (like e.g. a resonance transition
form factor extracted from pion electroproduction experiments) is well described in terms of quark degrees
of freedom at large momentum transfer, say $Q^2$ a few GeV$^2$, but then deviations between the model/theory show up when one
approaches the photon point at $Q^2 =0$. This is often cured by adding the pion cloud contribution from
some hadronic model, see e.g. some examples in the opening talk by Burkert~\cite{Burkert}.
I would like to take the opportunity to issue a warning here. While the notion of the ``pion cloud'' is
fairly intuitive, it is rather difficult to assign to it a quantitative measure. So let us look at that in more detail.

The so-called pion cloud is best analyzed in chiral perturbation theory.
There is one very intriguing structural aspect generated by the (almost) massless
Goldstone bosons of QCD, namely the so-called {\em chiral logarithms (logs)} or chiral singularities.
Pions couple to themselves and to matter fields like the nucleons, thus generating a cloud that
is Yukawa-suppressed at large distances as $\exp(-M_\pi r)/r$ for finite pion mass. In the chiral
limit of vanishing quark and thus pion masses $M_\pi\to 0$, this Yukawa tail turns into
a long-range Coulomb-like form. Consequently, S-matrix elements or transition
currents can diverge in this limit. Famous examples are the pion vector radius or the nucleon 
isovector radius that scale as $\log(M_\pi)$, see e.g. Ref.~\cite{Pagels:1974se}, or the nucleon electromagnetic 
polarizabilities that are proportional to $1/M_\pi$~\cite{Bernard:1991rq}. This is not a disaster but rather a 
natural consequence of having massless degrees of freedom, and it can serve
as an important check for other calculational schemes, like e.g. the lattice
formulation of QCD. Also, any model that is supposed to describe the pion or
the nucleon structure at low energies should obey such constraints. Note also that
such singularities are generated by loop graphs, so that in general there will
also be a pion-mass-independent contribution from a counterterm (contact interaction)
at the same order of the calculation. A typical example is the isovector charge radius of the pion,
or any other observable that displays a chiral log. Since the argument of a
log must be a number, it really depends on $M_\pi/\mu$, with $\mu$ some
regularization scale. Similarly, the corresponding LEC that appears must
also depend on $\mu$, thus making the observable scale-independent.  This,
however, also means that it is in general not possible to assign 
a definite value to the pion cloud
contribution of an observable since shifts in the regularization scale allow one
to shuffle strength from the long-range pion contribution to the shorter
ranged contact term part. I refer the reader to Ref.~\cite{s1Bernard:1998gv},
where the isovector charge radius of the proton is analyzed at one-loop order. It is
shown that a modest change in the regularization scale from $\mu =0.8$~GeV to
$\mu =1$~GeV (using conventional dimensional regularization) even leads to a
sign change in the counterterm contribution. This explicitly shows that there is,
of course, a contribution of the chiral (pion) physics to a given observable, but
only the observable quantity (here: the isovector charge radius) is independent
of the regularization scale. A more detailed discussion of this and related
issues can be found in Refs.~\cite{Meissner:2007tp,Henley:2013jga}.

\section{Summary and outlook}

The lessons learned and the take-home messages from this talk are:
\begin{itemize}
\item Resonances are defined as poles in the complex energy plane.
  To calculate their complex properties, one must locate the corresponding pole
  and then derive the resonance characteristics from suitably defined
  Laurent expansions around the pole position. Any other approach will in most
  cases lead to  imprecise or even wrong results. Be aware of {\em how}
  many entries in the PDG tables have been obtained!
\item The QCD spectrum is more than a collection of quark model states. The
  quark model should not be considered as a faithful representation of the
  QCD spectrum but rather as a simple approximation to a part of it. Note
  that the approximations underlying the conventional quark model are clearly
  better justifed for the heavy than for the light quarks.
\item Structure formation in QCD ties nuclear and hadron physics together.
  This is most clearly seen through the appearance of molecular-type
  structures both in the nuclear as well as in the hadronic landscape.
  This was realized early by some researchers, see e.g.~\cite{Voloshin:1976ap,Tornqvist:1993ng}.
\item Lattice QCD is making progress in addressing complex resonance properties,
  even finding poles deep inside the complex plane, like for the lowest resonance
  in QCD, the $f_0(500)$, see e.g.~\cite{Briceno:2016mjc,Guo:2018zss}.
  However, the extrapolation into the complex plane must respect chiral symmetry.
  Otherwise, one can easily miss a possible two-pole structure which was first observed
  for the $\Lambda(1405)$ but appears to play an even bigger role for states
  involving charm and bottom quarks. 
\item EFTs are of utmost importance in pushing this program forward. This has
  become apparent in the discussion of finding poles in the complex plane.
  The explicit calculations of finite-volume effects in hadron decays
  are another important playground for EFTs and this program is pursued vigorously, see
  e.g. the talk by Pang~\cite{Pang}.
\end{itemize}

\section*{Acknowledgments}

I thank all my collaborators for sharing their insights on the various topics discussed here.
Special thanks to Maxim Mai for supplying Fig.~\ref{fig:L1405}, and to
Feng-Kun Guo,  Bernard Metsch and Akaki
Rusetsky for a careful reading of the manuscript.
This work is supported in part by  the DFG (Grant No. TRR110)
and the NSFC (Grant No. 11621131001) through the funds provided
to the Sino-German CRC 110 ``Symmetries and the Emergence of
Structure in QCD",  by the Chinese  Academy of Sciences (CAS) President's International 
Fellowship Initiative (PIFI) (grant no. 2018DM0034)  and by VolkswagenStiftung (grant no. 93562).
Computational resources for this project were provided 
by the J\"{u}lich Supercomputing Centre (JSC) at the Forschungszentrum 
J\"{u}lich  and by RWTH Aachen.

%

\end{document}